\documentclass[aps,prl,twocolumn,showpacs,amsmath,amssymb,superscriptaddress]{revtex4-2}
\usepackage{graphicx}
\usepackage{CJK}
\usepackage{color}
\usepackage{appendix}
\usepackage{bm}
\usepackage[colorlinks,bookmarks=false,citecolor=blue,linkcolor=red,urlcolor=blue]{hyperref}
\usepackage{multirow}
\usepackage{ulem}
\usepackage{tabularx}
\usepackage{booktabs}
\usepackage[nounderscore]{syntax}

\newcommand{\be}{\begin{equation}}
\newcommand{\ee}{\end{equation}}

\newcommand\GWA[1]{{\textsf{\footnotesize{\color{red}[GWA: #1]}}}}
\begin{document}
\title{Deconfined quantum critical point in a dissipative spin-1/2 chain}
\author{Longye Lu}
 \affiliation{School of Physics and Astronomy, Beijing Normal University, and Key Laboratory of Multiscale Spin Physics (Beijing Normal University), Ministry of Education, Beijing 100875, China}
\author{Shifeng Cui}
\affiliation{Shandong Engineering Research Center of Aeronautical Materials and Devices, College of Aeronautical Engineering, Shandong University of Aeronautics, Binzhou 256603, China}
\author{Wenan Guo}
\email{waguo@bnu.edu.cn}
\affiliation{School of Physics and Astronomy, Beijing Normal University, and Key Laboratory of Multiscale Spin Physics (Beijing Normal University), Ministry of Education, Beijing 100875, China}
\date{\today}

\begin{abstract}
Open quantum spin systems offer a previously unexplored route to realizing deconfined quantum criticality.
We consider a spin-1/2 $J$-$Q_3$ chain, consisting of an antiferromagnetic (AFM) Heisenberg exchange and a competing multi-spin interaction favoring a valence-bond solid (VBS) state, with each spin component coupled to a bosonic bath. Using non-Abelian bosonization and renormalization-group (RG) analysis, combined with large-scale quantum Monte Carlo (QMC) simulations, we determine the phase diagram and the associated phase transitions of the model. We show that strong dissipation stabilizes an AFM phase for sub-Ohmic, Ohmic, and super-Ohmic baths.
Continuous  AFM-VBS transitions at finite dissipation are found upon increasing the multi-spin interaction in both the sub-Ohmic and Ohmic regimes. Critical properties are obtained through perturbative RG analysis and QMC simulations. In the Ohmic case, the critical point features spinon deconfinement and emergent O(4) symmetry.  In the sub-Ohmic regime, the transition may also involve spinon deconfinement, provided that spinons remain deconfined in the dissipative VBS phase. In addition, in the super-Ohmic regime, we propose a transition from AFM phase to a quasi-long-range ordered phase. 
\end{abstract}
\maketitle

\textit{Introduction}---
The deconfined quantum critical point (DQCP) \cite{Senthil_2004, Senthil_2004_2} is a generic scenario for continuous phase transition from the antiferromagnetic state (AFM)  to valence-bond solid (VBS) that beyond the paradigm of Landau-Ginzburg-Wilson. Its defining feature is the deconfinement of spinons at criticality, while the remain confined in both ordered phases \cite{Levin-Senthil2004}. 
Over the past two decades, the DQCP has been extensively investigated in 
two-dimensional(2D) quantum spin, three-dimensional classical loop, and fermionic models (see, e.g., \cite{senthil_review} and references therein).
Although many signatures of the DQCP have been observed \cite{shaoGuoSandvik-Science}, 
a growing consensus is that the putative DQCPs realized in existing microscopic models are ultimately weakly first-order rather than genuinely continuous \cite{wangchong-prx2017, Nahum-PRX2015, Nakayama-PRL, dengliu2024}. A multicritical variant of the DQCP 
has, however, been proposed recently \cite{takahashi-multi, chester-su-prl2024, Li-Shen-bootstrapcone}.

An alternative route to realizing DQCP is to introduce nonlocal interactions in one-dimensional (1D) systems \cite{yangsbJQchain, JianCM1DDQC, XuYCDQCnonlocal}.  By introduing long-range instantaneous interacions\cite{yangsbJQchain}, or, by coupling the system to a critical bulk\cite{XuYCDQCnonlocal, JianCM1DDQC, Song-Zhang-boundary2025, ding_deconfinedboundary}, 1D analogues of the DQCP have been proposed and studied both numerically and theoretically. 
Open quantum spin systems \cite{Weiss_book} offer a previously unexplored route to realizing DQCP. The  environment can be modeled by bosonic baths, whose integration  naturally results in retarded nonlocal interactions of spin degrees of freedom  \cite{Leggett-RMP}. In this letter, we consider a 1D spin-1/2 system, namely the $J$-$Q_3$ chain\cite{TangYing1dJQ3, Kedar1dJQ3}, with each spin component coupled to an independent bosonic bath, which is classified as sub-Ohmic ($s < 1$), Ohmic ($s=1$), and super-Ohmic ($s > 1$) according to the power-law exponent $s$ of the spectral density, $J(\omega)\sim \omega^s$ \cite{Weiss_book}.
The resulting retarded nonlocal interaction decays as $\tau^{-(1+s)}$, where $\tau$ denotes the imaginary time. 
Such interactions in the Ohmic case have been found to stabilize the AFM order\cite{DissipationInducedOrder}, which is forbidden in the Heisenberg chain by the Mermin-Wagner theorem \cite{Mermin1966, Hohenberg} . The competition between dissipation, which tends to stabilize AFM order, and multi-spin interaction, which favors a VBS state, gives rise to rich quantum critical behavior.

Using non-Abelian bosonization and renormalization group (RG) analysis, 
we reveal the phase diagram and associated phase transitions of the dissipative system:
When coupled to the sub-Ohmic bath, the system undergoes a continuous AFM-VBS transition,  
which is controlled by a fixed point (FP) with dynamical exponent $z>1$, and the scaling dimensions of the N\'eel and VBS order parameters are $\Delta_N>1/2$ and $\Delta_D<1/2$, respectively. 
The transition persists up to the Ohmic case, where the FP merges into the SU(2)$_1$ CFT with $\Delta_N=\Delta_D=1/2$ and $z=1$. Consequently, the AFM-VBS critical point  
hosts deconfined spinons and an emergent $\mathrm{O}(4)$ symmetry--well established characteristics of the isolated Heisenberg chain--thus corresponding to a DQCP with emergent symmetry\cite{Nahum_SO5}. In particular, spinons are confined in the AFM state, making this transition distinct from the 1D analogues of the DQCP between two Ising-type ordered phases\cite{Motrunich-1d-dqc, Hikihar-1d-dqc, XiangTao-1D-DQC,weber2025competingdiracmassesdimension}. 
The AFM-VBS transition in the sub-Ohmic regime may also involve the deconfinement of spinons upon their entering the VBS state from the AFM state, and can thus be considered an analogue of the 2D DQCP, albeit without emergent $\mathrm{O}(4)$ symmetry, provided that the deconfinement persists in the VBS state as in the isolated case \cite{TangYing1dJQ3, TangYing-1dspinonsPRB}.  
Upon entering the super-Ohmic regime, another FP 
with $z > 1, \Delta_N < 1/2$, and $\Delta_D > 1/2$ emerges, describing a transition from the AFM phase to the QLRO phase controlled by the SU(2)$_1$ CFT.    
Subsequently, we present QMC simulations of our model that corroborate our theoretical findings.

\textit{Model and RG analysis}---
The spin-1/2 $J$-$Q_3$ chain is described by\cite{TangYing1dJQ3, Kedar1dJQ3}
\begin{equation}
H_{\rm s}=-J\sum_{i=1}^N  C_{i, i+1}-Q \sum_i C_{i, i+1} C_{i+2, i+3} C_{i+4, i+5},
\end{equation}
where $C_{ij}=1/4-\mathbf{S}_i\cdot \mathbf{S}_j$ is the singlet projector.  The Heisenberg exchange coupling $J=1$ sets the energy unit. The six-spin coupling $Q\ge 0$. The model reduces to the standard Heisenberg model when $Q=0$. Upon increasing $Q$, the ground state changes from a critical phase to a VBS at $Q_{c,0}\approx 0.1645$ \cite{TangYing1dJQ3, Kedar1dJQ3}.     

The effects of dissipation are studied by introducing  an independent bosonic bath that is coupled to each spin component $S_i^\alpha$, and the total Hamiltonian reads $ H=H_{\text s}+H_{\rm sb}$, with
\begin{equation}\label{h_sb}
   H_{\rm sb}=\sum_{i q} \omega_q \mathbf{a}_{i q}^{\dagger} \cdot \mathbf{a}_{i q}+\sum_{i q} g_q (\mathbf{a}_{i q}^{\dagger}+\mathbf{a}_{i q}) \cdot \mathbf{S}_i,
\end{equation}
where $\mathbf{a}_{iq}$ and $\mathbf{a}^\dagger_{iq}$ are three-component vectors of bosonic annihilation and creation operators, respectively, at site $i$ for a mode $q$ with frequency $\omega_q$, and $g_q$ is the spin-boson coupling. 
The model has a global SO(3) symmetry generated by the total angular momentum of the total system \cite{DissipationInducedOrder}. 

Integrating out the bosonic degrees of freedom exactly yields an effective retarded interaction of the spins 
\begin{equation}
    H_{\rm ret}=-\iint_0^\beta d\tau d\tau'\sum_{i}K(\tau-\tau')\mathbf{S}_i(\tau)\cdot \mathbf{S}_i(\tau'),
\end{equation}
where $K(\tau-\tau')$ is the propagator:
\begin{equation}
   K(\tau) =\frac{1}{\pi}\int_0^{\omega_c}d\omega J(\omega)\frac{e^{-\omega\tau}}{1-e^{-\beta \omega}}\sim \frac{\alpha}{|\tau|^{1+s}},
\end{equation}
with  $J(\omega)=2\pi\alpha \omega_c^{1-s}\omega^s$ being the spectral density of the bath,
$\alpha$ is the dimensionless coupling constant, $\omega_c$ is the frequency cutoff, and $\beta$ is the inverse temperature. The partition function is then determined by the spin subsystem with ${\cal H} =H_{\rm s}+H_{\rm ret}$.

We now adopt non-Abelian bosonization to approach the model ${\cal H}$ 
featuring dissipation via three independent baths, in contrast to the Abelian bosonization applied to one-bath dissipative spin chains\cite{bosonization_dissipation, SciPostPhys_bosonization}.

The bosonized form of the Hamiltonian is written as $H=H[\mathrm{SU}(2)_1]+
\lambda \int dxd\tau \vec{J_L}\cdot \vec{J_R}+H_\phi+ g a^{(s-1)/2} \int dx d\tau \vec{\phi}\cdot \vec{n}$, 
  where  $\vec{J}_{L, R}$ are the left and right chiral modes generating the SU(2)$_{L, R}$ symmetries, 
      $H[\mathrm{SU}(2)_1] =\frac{1}{6\pi}\int dxd\tau \left[\vec{J}_L\cdot \vec{J}_L+\vec{J}_R\cdot \vec{J}_R\right]$,
   $\lambda$ is the back-scattering interaction, 
   $H_\phi=\int dxd\tau  dx'd\tau' \phi^i(x,\tau)C_{ij}^{-1}(x,\tau;x',\tau')\phi^j(x',\tau')$ with $ C_{ij}(x,\tau;0,0)=\langle\phi^i(x,\tau)\phi^j(0,0)\rangle=\frac{\delta_{ij}\delta(x)}{|\tau|^{1+s}}$,
   is the Hamiltonian of generalized free field $\vec{\phi}$; $\vec{n}$ is the N\'eel order parameter,
    $g$ generates the imaginary-time long-range interaction induced by dissipation, and $a$ is the real space cutoff. The beta functions are obtained by perturbing the $\mathrm{SU}(2)_1$ CFT: 
    \begin{align}\label{beta_functions}
    \beta_g=\frac{dg}{dl}&=\frac{\varepsilon}{2}g-\frac{\pi}{2}\lambda g,\nonumber\\
    \beta_\lambda=\frac{d\lambda}{dl}&=2\pi\lambda^2-\frac{1}{2}g^2,
\end{align}
where $\varepsilon\equiv 1-s$.
Details of the derivation can be found in the Supplemental Materials\cite{supplemental}.

\begin{figure}[htpb]
	\includegraphics[width=0.45\textwidth]{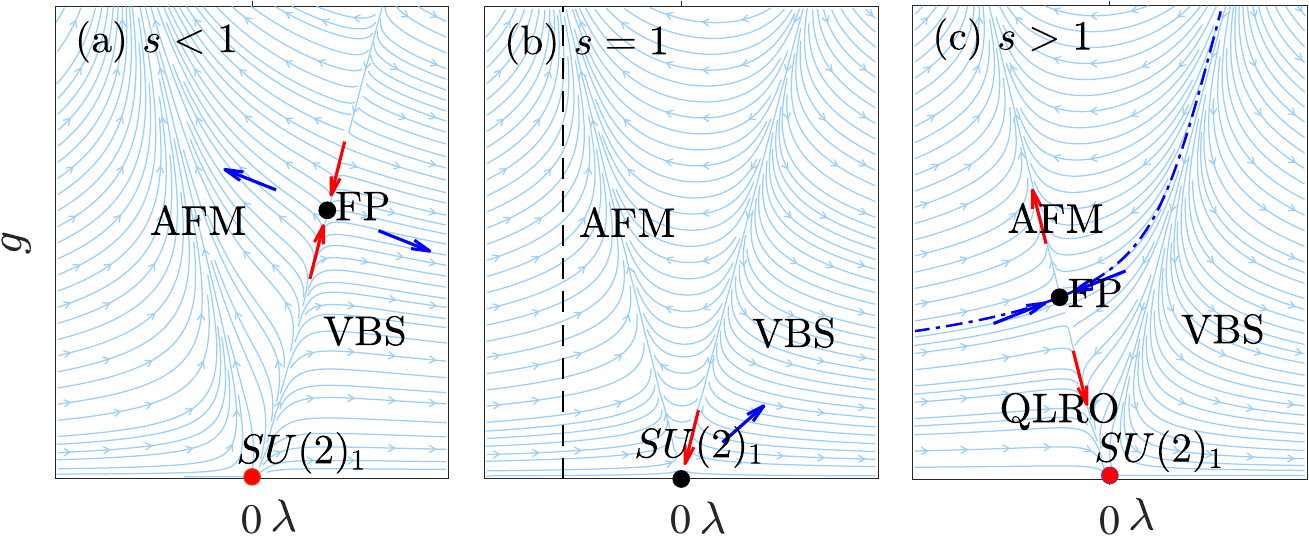}
	\caption{ (a) -(c) The RG flow diagram of beta function for $s<1$, $s=1$, and $s>1$, respectively. The short arrows near the FP are the eigen-directions. The black dashed line indicate the RG starting points of dissipative Heisenberg chain. The blue dot-dashed line indicates the phase boundary of QLRO phase and AFM phase.
    }
	\label{Fig:RG_flow}
\end{figure}
The resulting RG phase diagrams are shown in Fig.~\ref{Fig:RG_flow} for the sub-Ohmic, Ohmic, and super-Ohmic cases.
In all cases, a strong-dissipation FP appears at $(\lambda, g)\to (-\infty,+\infty)$, representing an AFM phase. At this FP, strong backscattering locks the Wess-Zumino-Witten (WZW) primary field, $\mathbf{g} = I_{2\times 2} + i \vec{n}\cdot\vec{\sigma}$, to $i \vec{n}\cdot\vec{\sigma}$, where $\vec{\sigma}$ are Pauli operators. 
The large dissipation strength strongly damps the quantum dynamics along the imaginary-time direction, enforcing $\partial_\tau \vec{n} \approx 0$. Consequently, the WZW action reduces to a 1D classical nonlinear $\sigma$-model, $S_{\mathrm{WZW}} \sim \beta \int dx \, (\partial_x \vec{n})^2$, 
which is dominated by the uniform classical configuration $\partial_x \vec{n} = 0$ at zero temperature, leading to long-range order.
This reveals how dissipation stabilizes AFM order, consistent with 
the Ohmic dissipative Heisenberg chain\cite{DissipationInducedOrder} and applicable to sub-Ohmic and super-Ohmic cases as well.

For the sub-Ohmic dissipation ($s < 1$), as the backscattering interaction increases at a finite $g$, the system undergoes a continuous AFM--VBS transition, and 
ultimately flows under RG toward the VBS phase, where the $\mathbb{Z}_2$ lattice translational symmetry is spontaneously broken, as shown in Fig. \ref{Fig:RG_flow}(a). 
The transition is governed by a non-trivial FP at
$(\lambda^*,g^*)=(\varepsilon/\pi,2\varepsilon/\sqrt{\pi})$.
The two eigen-directions consist of one relevant direction, which flows toward the AFM/VBS phases, and an irrelevant direction, which spans the critical surface. 
The Critical properties are obtained as $\Delta_N = 1/2 + \varepsilon/2$, $\Delta_D = 1/2 - 3\varepsilon/2$, $\nu = 1/[(2+\sqrt{6})\varepsilon]$ to order $\varepsilon$, and $z = 1 + 3\varepsilon^2/\pi^2$ to order $\varepsilon^2$ \cite{supplemental, antunes2026lifshitzcriticalpointsmeet}.
In the Ohmic case ($\epsilon \to 0$), the FP merges into 
the $\mathrm{SU}(2)_1$ CFT FP with $\Delta_N=\Delta_D=1/2$ and $z=1$, as shown in Fig.\ref{Fig:RG_flow}(b). 

In the super-Ohmic ($s>1$) case, power-counting at the $\mathrm{SU(2)_1}$ CFT point suggests that dissipation is irrelevant. However,  Eq.~(\ref{beta_functions}) 
reveals a continuous phase transition governed by another FP $(\tilde{\lambda^*},\tilde{g^*})=(\varepsilon/\pi,-2\varepsilon/\sqrt{\pi})$, separating the AFM phase from a QLRO phase described by the $\mathrm{SU(2)_1}$ CFT. 
This FP is mathematically `dual' to the FP $(\lambda^*,g^*)$, with critical properties $\Delta_N =1/2+\varepsilon/2$, $ \Delta_D =1/2-3\varepsilon/2 $, $\nu= -1/[(2+\sqrt{6})\varepsilon]$, and $z=1+3\varepsilon^2/\pi^2$.  
With increasing dissipation, the QLRO phase becomes very narrow, so the transition could be misread as a DQCP governed by the $\mathrm{SU}(2)_1$ CFT. Our subsequent numerical simulations are restricted to the Ohmic and sub-Ohmic cases only.

\textit{Numerical results--}Now we present  QMC simulation results for the dissipative $J$-$Q_3$ chain. 
The model ${\cal H}=H_{\rm s}+H_{\rm ret}$ is simulated using a recently developed QMC method 
that incorporates a wormhole algorithm for retarded interactions\cite{Weber-wormhole, Weber-QMC4ret}. This method is based on the stochastic series expansion with loop updates~\cite{Sandviksusc1991, Sandvik1999}. We use periodic boundary conditions and take the bosonic frequency cut-off $\omega_c=10 J$. 
 To access ground-state physics, we carefully choose different inverse temperature 
 for different $s$, see \cite{supplemental} for details. Typically, several $10^6$ Monte Carlo steps (MCs) are used to reach equilibrium, after which measurements are taken over 100 bins of $10^4$ MCs each.

We probe the AFM phase and its boundaries using the Binder cumulant based on the staggered magnetization $m_z=\frac{1}{L}\sum_i(-1)^i S^z_i$, 
\begin{equation}
	U_m=\frac{3}{2} \left(1 -\frac{1}{2}\frac{\langle m_z^4\rangle}{\langle m_z^2 \rangle^2}\right),
\end{equation}
or, alternatively, using the correlation ratio
\begin{equation}
    R=1-\frac{S(\pi+\delta q)}{S(\pi)},
\end{equation}
where $S(q)=\sum_{ij}e^{iq(i-j)}\langle S_i^zS_j^z\rangle/L$ is the structure factor and $q=\pi$ is the AFM ordering momentum with resolution $\delta q=2\pi/L$. $R$ is proportional to $(\xi/L)^2$ with $\xi$ being the correlation length. $R$ and $U_m$ converge to 1 in the AFM ordered phase, while to 0 in the disordered phase. Both become size-independent at the critical point. 

The VBS phase and its boundaries are studied using the Binder cumulant $U_D$ associated with the VBS order parameter $\mathcal{D}=\frac{1}{L}\sum_i(-1)^iS^z_iS^z_{i+1}$:
\begin{equation}
U_D=\frac{3}{2}\left(1-\frac{1}{2} \frac{\langle \mathcal{D}^4\rangle}{\langle \mathcal{D}^2\rangle^2}\right).
\end{equation}
In the VBS phase, $U_D$ converges to 1 as $L\to \infty$, whereas it vanishes in the non-VBS phase and becomes size-independent at the critical point.
 
\begin{figure*}[htbp]
	\centering	
    \includegraphics[width=0.3\textwidth]{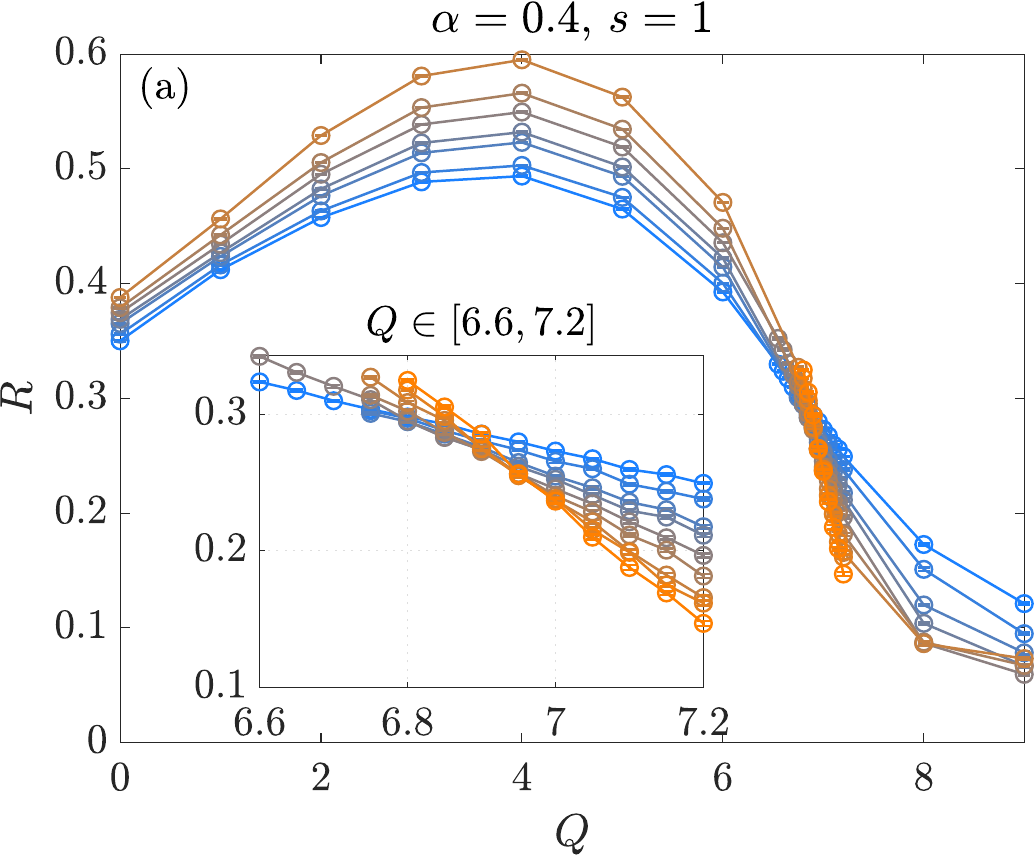}
    \includegraphics[width=0.3\textwidth]{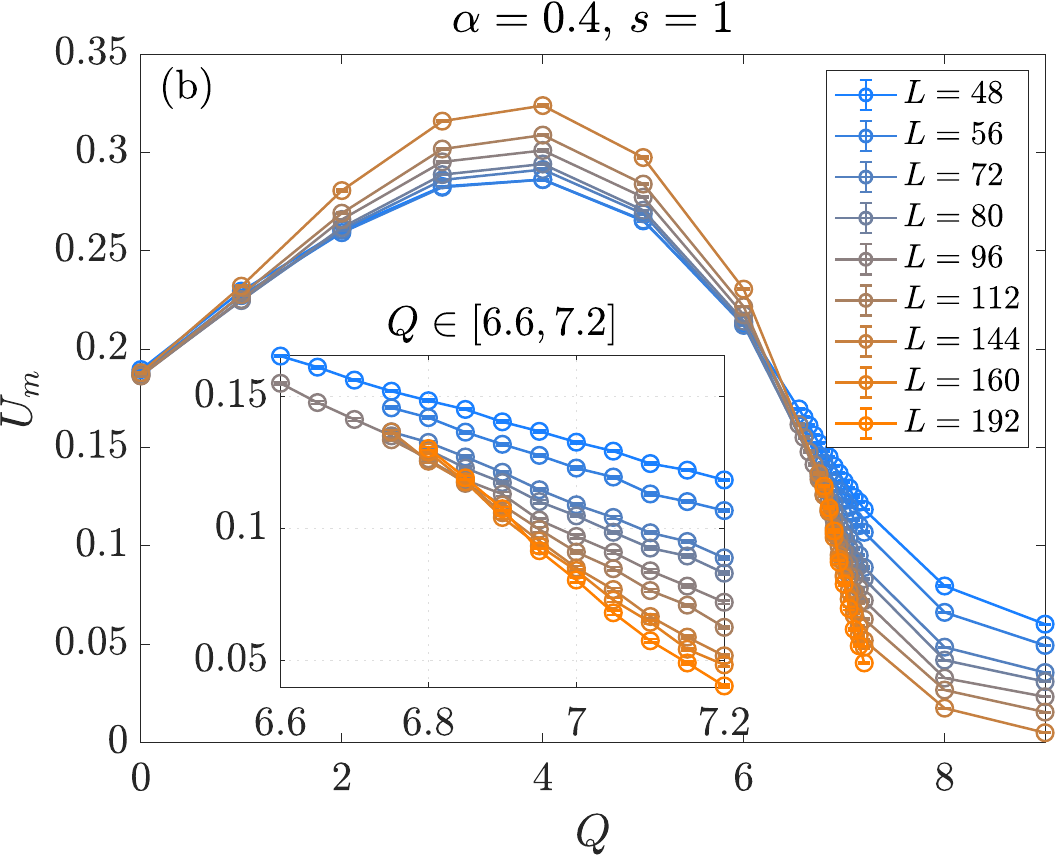}
	\includegraphics[width=0.3\textwidth]{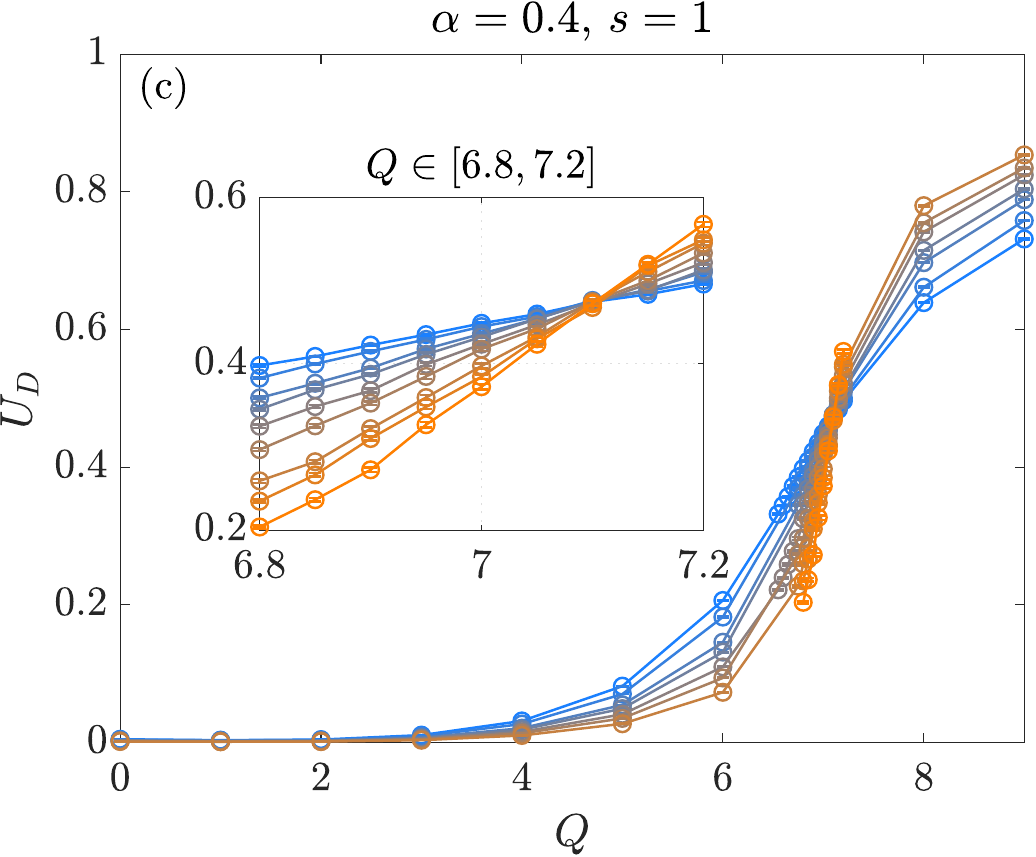}
    \caption{ (a)$R$, (b)$U_m$, and (c) $U_D$ versus $Q$ for different system sizes at coupling $\alpha=0.4$ for the Ohmic dissipation.
     }
    \label{Fig:Binder}
    \end{figure*}

Figure \ref{Fig:Binder}(a) and (b) show $R$ and $U_m$ versus $Q$ for various system sizes in the Ohmic case. 
At $Q=0$, $R(L)$ exhibits a slow convergence to 1, in agreement with \cite{DissipationInducedOrder}. 
Peaks around $Q=4$ grow with increasing system size, indicating AFM order and suggesting that it is enhanced by the $Q$ term, as explained by the RG flow and spin-wave analysis\cite{supplemental}. 
Further increasing $Q$ drives the system out of the AFM phase, as expected. 
Figure \ref{Fig:Binder}(c) plots $U_D$ versus $Q$ for various system sizes. 
For $Q \gtrsim 7$, $U_D$ converges to 1 as $L \to \infty$, indicating VBS order.
The crossings of the $U_m(Q)$, $R(Q)$, and $U_D(Q)$ for different sizes occur at roughly the same $Q$, suggesting a direct AFM-VBS transition at $Q_c$. 
The absence of a negative peak in the Binder cumulants rules out a conventional first-order transition \cite{Vollmayr}. 
Similar finite-size behaviors of $U_m, R$, and $U_D$ are observed in the sub-Ohmic case, e.g., $s=0.95, 0.85$\cite{supplemental}. The fact that $Q_c$ is significantly larger than $Q_{c,0}$  agrees well with the RG phase diagram Fig. \ref{Fig:RG_flow}.

We adopt the standard ($L, 2L)$ crossing analysis; see, e.g., the Supplemental Material of \cite{shaoGuoSandvik-Science}, 
to show that the crossings converge to the same value of $Q$. 
Near a critical point with distance $\delta=Q-Q_c$, a dimensionless quantity $A$, such as $U_m, R,$ and $U_D$, has the finite-size scaling form $A(\delta, L)=A(\delta L^{1/\nu}, uL^{-\omega})$, where $\nu$ denotes the correlation length exponent, $u$ represents irrelevant fields and $\omega>0$ the effective irrelevant exponent associated with $u$.  
The crossing point $Q_c^{A}(L)$ satisfying $A(Q,L)=A(Q,2L)$, 
converges to the critical point $Q_c$ in the following way
\begin{equation}
    Q_c^{A}(L)-Q_c \sim  L^{-1/\nu -\omega}.
   \label{qcfss}
\end{equation}

\begin{figure}[htbp]
    	\centering
	\includegraphics[width=0.22\textwidth]{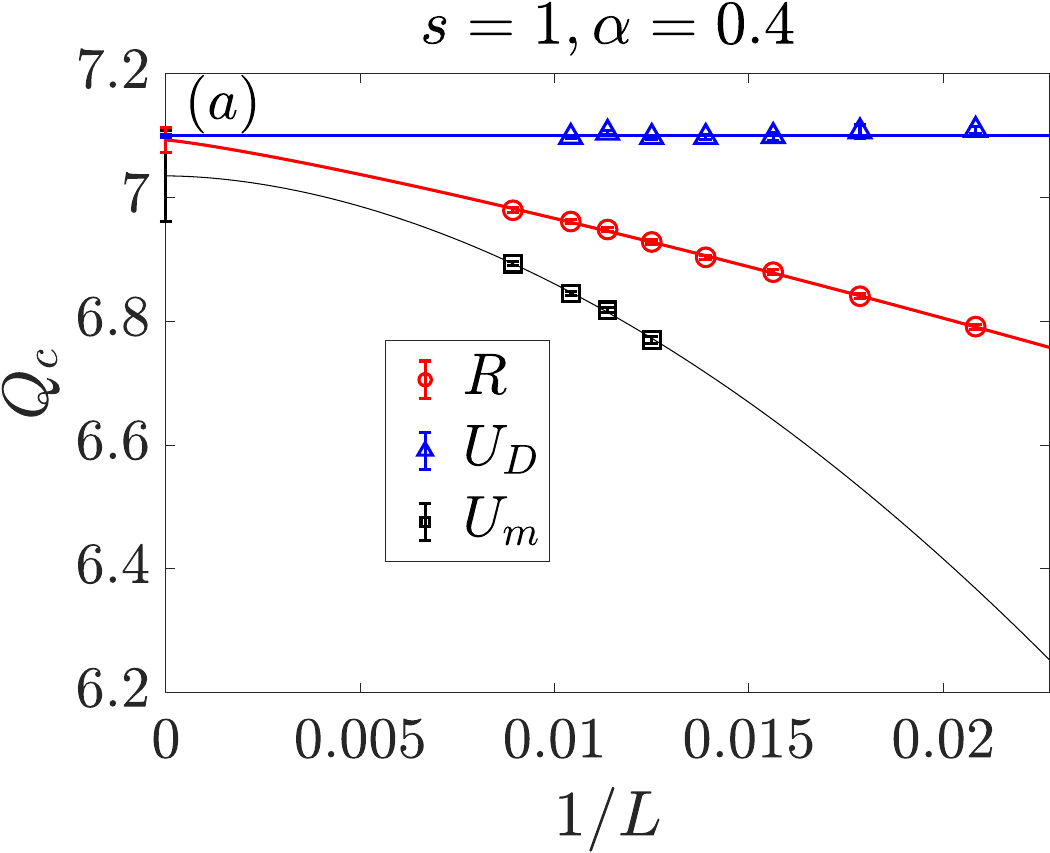}
    	\includegraphics[width=0.22\textwidth]{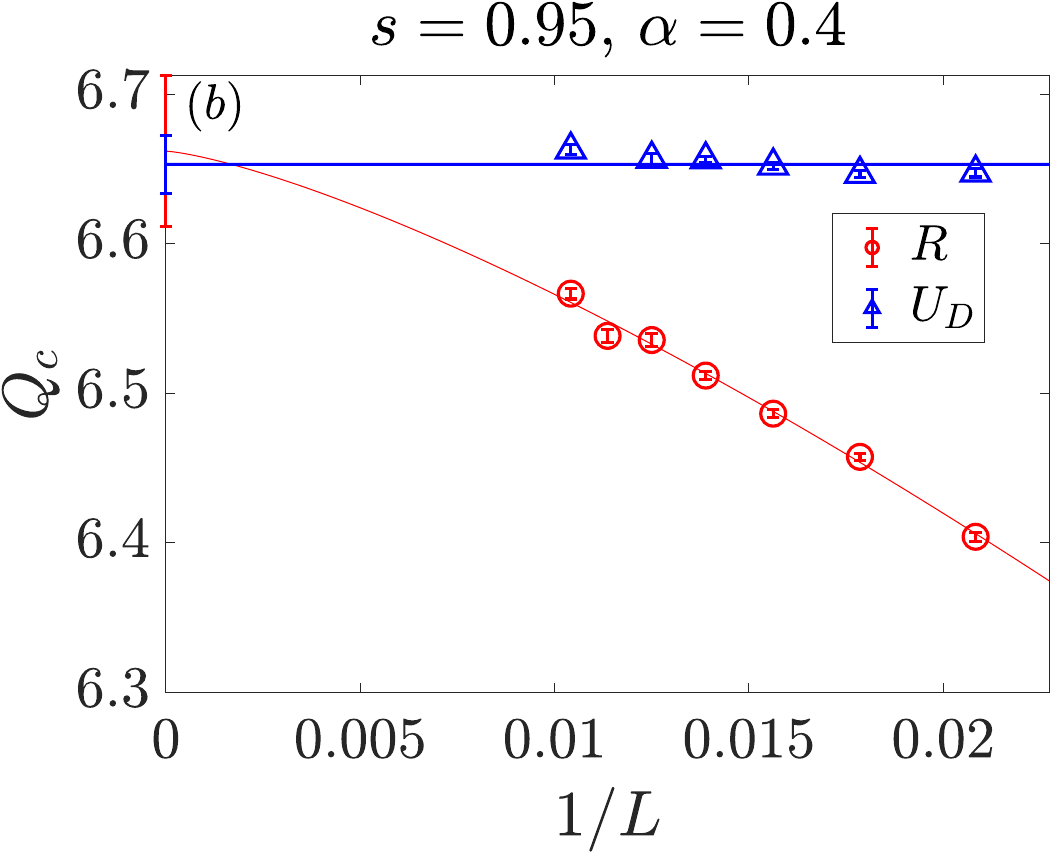}
	\caption{Crossing-point analysis for the Ohmic (a) and sub-Ohmic ($s=0.95$) (b) cases with $\alpha=0.4$. The symbols mark the crossings of the $U_m, R$, and $U_D$ curves between $L$ and $2L$. For $s=0.95$, the crossings of $U_m$ are omitted because of poor convergence resulting from strong corrections to scaling. 
    Solid lines are power-law fits to crossings of $U_m$ and $R$, and cubic polynomial fit to crossings of $U_D$. 
    The extrapolated $Q_c$ are indicated at $1/L=0$. 
    } 
	\label{Fig:Qc(L)}
\end{figure}

The crossings $Q_c^A(L)$ versus $1/L$ for $A=U_m, R$, and $U_D$ are displayed in Fig. \ref{Fig:Qc(L)} for the Ohmic and sub-Ohmic ($s=0.95$) cases. Power-law fits of Eq. (\ref{qcfss}) for $A=U_m$ and $R$ are performed. For unknown reasons, the irrelevant field for $A=U_D$
is very small, making crossings of $U_D$ curves difficult to fit with Eq.(\ref{qcfss}); cubic polynomials are used instead.
For both cases, the $Q_c$ values extracted from the three quantities agree within one error bar, supporting a direct N\'eel-VBS transition. Our best estimates are $Q_c=7.100(2)$ for $s=1$ and $Q_c=6.65(2)$ for $s=0.95$. The results are listed in Tab. \ref{tab:qcs}. 

\begin{table*}[htbp]
\centering
\caption{Critical properties of the AFM-VBS transition at different $s$ with $\alpha=0.4$. $z(\xi_\tau)$ is extracted from $\xi_\tau$; $\Delta_N(C)$ and $ \Delta_N(m_s^2)$ from $C(L/2)$ and $m_s^2(L)$; $\Delta_D(D^2)$ and $ \Delta_D(C_D)$ from $D^2(L)$ and $C_D(L/2)$.
$z(T), \Delta_N(T)$, and $\Delta_D(T)$ denote RG results. }
\label{tab:qcs}
\begin{tabular}{cc cc ccc ccc}
\toprule
$s$ & $Q_c$ 
   & $z(\xi_\tau)$ & $z(T)$ & $\Delta_N(C)$ & $\Delta_N(m_s^2)$ & $\Delta_N(T)$ & $\Delta_D(D^2)$ & $\Delta_D(C_D)$ & $\Delta_D(T)$ \\
\midrule
1.00 & 7.100(2) & 1.02(2) & 1     & 0.499(3)  & 0.5018(4) & 0.5    & 0.50(1)  & 0.502(2) & 0.499(1)   \\
0.95 & 6.65(2) & 1.12(1) & 1.002 & 0.53(2)  & 0.50(2) & 0.525 & 0.259(5) & 0.25(1) & 0.425 \\
0.85 & 5.963(6) & 1.97(7) & 1.02  & 0.46(2) & 0.443(4) & 0.575  & 0.248(2) & 0.240(5) & 0.275 \\
\bottomrule
\end{tabular}
\end{table*}

To obtain the scaling dimension $\Delta_N$, we study the scaling of the longest-distance spin correlation at the estimated $Q_c$ 
\begin{equation}
    C(L/2)\sim 
    \begin{cases}
L^{-2\Delta_N}, & s<1,\\[2mm]
(\ln L)^{q(C)}L^{-2\Delta_N}, & s=1.
\end{cases}
    \label{corrfss1}
\end{equation}
where $C(r)=(-1)^r\langle S_i^z S_j^z\rangle$ with $r=|i-j|$. 
The multiplicative logarithmic correction to the power law in the Ohmic case is predicted theoretically\cite{supplemental}.
Eq.(\ref{corrfss1}) also holds for
the squared AFM order parameter $m_s^2(L)=\langle m_z^2\rangle$ with $q(C)$ replaced by $q(m_s)$.

\begin{figure}[htbp]
    	\centering
	\includegraphics[width=0.4\textwidth]{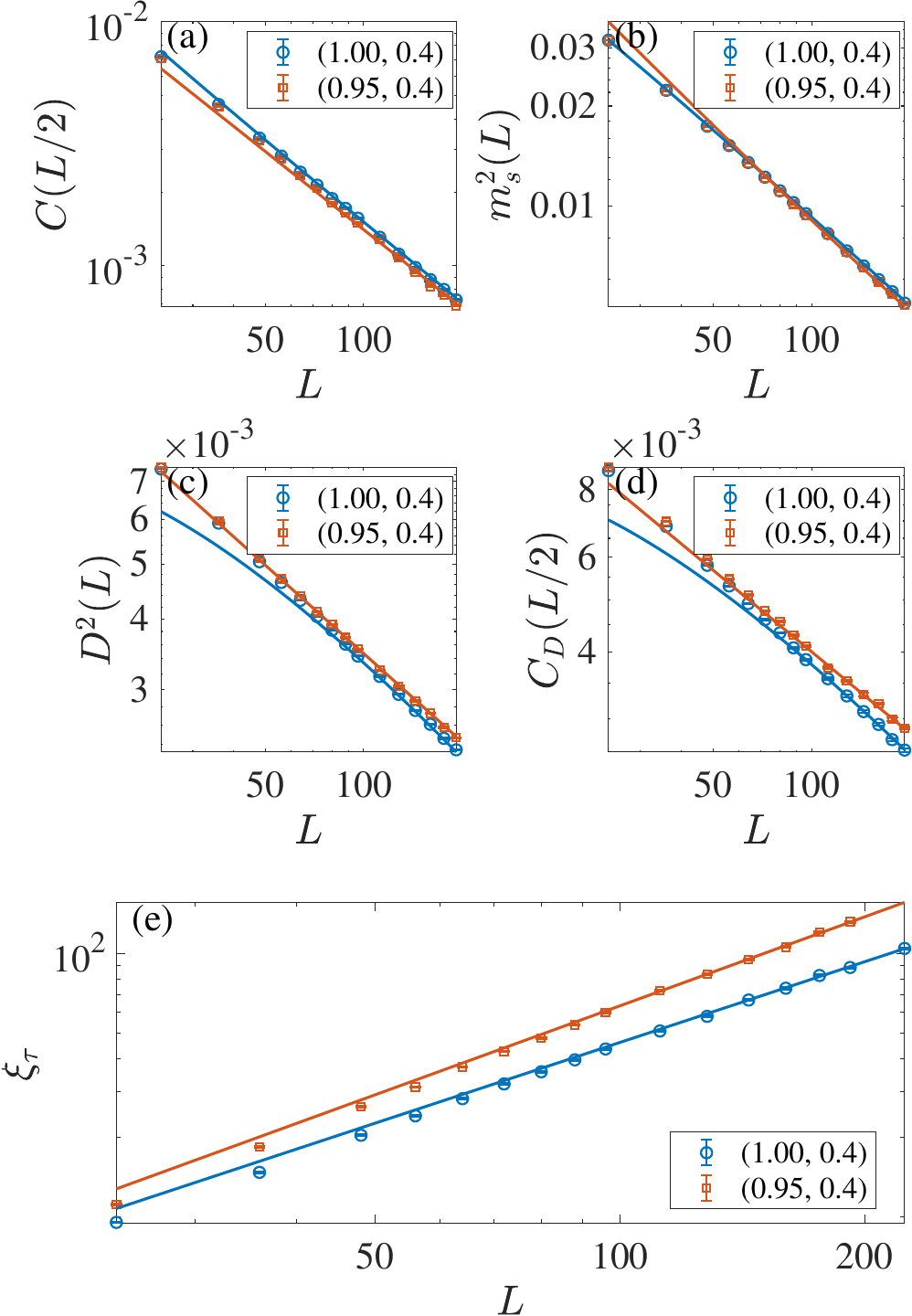}
    \centering
	\caption{Log-Log plots of $C(L/2)$ (a), $m_s^2(L)$(b), $D^2(L)$(c), $C_D(L/2)$(d), and $\xi_\tau$ (e)  at the AFM-VBS critical points for different $(s, \alpha)$. Solid lines are fits.} 
	\label{Fig:allcl}
\end{figure}

Figure \ref{Fig:allcl}(a) shows $C(L/2)$ at $\alpha=0.4$ for sub-Ohmic ($s=0.95$) and Ohmic ($s=1$) dissipation. A fit to Eq.(\ref{corrfss1}) yields $\Delta_N=0.53(2)$ for $s=0.95$. 
For $s=1$, we fit Eq.(\ref{corrfss1}) to the data in two steps: 
first fixing $\Delta_N=1/2$ to determine $q$, and then fixing the obtained $q$ to extract $\Delta_N$. The reliability of the results is guaranteed self-consistently. By gradually discarding the smallest system sizes, we eliminate systematic errors from subleading corrections to scaling. Our best estimates are $q(C)=-0.49$, $\Delta_N=0.499(3)$, 
which agree remarkably well with $\mathrm{SU}(2)_1$ CFT prediction $\Delta_N=1/2$. 

These results are further verified by fitting $m_s^2(L)$, as shown in Fig. \ref{Fig:allcl}(b). Consistent values of $\Delta_N$ for $s=0.95$ and $s=1$ are obtained. 
In particular, for $s=1$, we obtain $q(m_s^2)=0.53$, $\Delta_N=0.5018(4)$. 
The relation $q(m_s)\approx q(C)+1$ is a characteristic feature of the multiplicative logarithmic correction, since $m_s^2(L)$ is the spatial integral of $C(r)$. 
The best estimates of $\Delta_N$ are listed in Tab. \ref{tab:qcs}.

We extract the scaling dimension $\Delta_D$ from the scaling of the squared VBS order parameter $D^2=\langle \mathcal{D}^2 \rangle$  at $Q_c$: 
\begin{equation}
D^2(L) \sim
\begin{cases}
L^{-2\Delta_D}, & s<1,\\[2mm]
(\ln L)^{q(D)}L^{-2\Delta_D}, & s=1.
\end{cases}
\label{D2fss}
\end{equation}
The same scaling also apply to dimer correlation function
$C_D(r)=(-1)^r(C_B(r)-\frac{1}{2}(C_B(r+1)+C_B(r-1))$,
where $C_B(r)\equiv \langle (S_1^zS_2^z)(S_{r+1}^zS_{r+2}^z)\rangle$,
with $q(D)$ replaced by $q(C_D)$.
Figure \ref{Fig:allcl} (c) and (d) display $D^2(L)$ and $C_D(L/2)$ at $\alpha=0.4$ for $s=0.95$ and $s=1$, respectively. 
Fitting Eq.~(\ref{D2fss}) to these data, we find $\Delta_D$ (listed in Tab.~\ref{tab:qcs}) and $q(D)\sim 2.19, q(C_D) \sim 2.18$. Together with $q(C)\approx-0.5$, these values agree well with the $\mathrm{SU}(2)_1$ CFT relation $q(C_D)=-3q(C)$\cite{supplemental}, strongly supporting that the system flows back to the $\mathrm{SU}(2)_1$ CFT along the critical surface of the AFM-VBS FP in the Ohmic case.  

We now determine the dynamical critical exponent $z$ at AFM-VBS transitions
from the imaginary-time correlation length $\xi_\tau$ 
\cite{Matsumoto_PRB2001}, 
\begin{equation}
    \xi_\tau=\frac{\beta}{2\pi}\sqrt{\frac{S(\pi,0)}{S(\pi,\omega_1)}-1},
\end{equation}
rather than from the uniform susceptibility $\chi_u$ (for reasons see \cite{supplemental}).
Here $S(q,\omega)$ is the imaginary-time dynamical structure factor,
$S(q,\omega)=\frac{1}{L}\sum_{mn}\int_0^\beta d\tau e^{-iq (m-n)-i\omega \tau} \langle S_m^z(\tau)S_n^z(0)\rangle$, 
where $q$ is the wave vector and $\omega_1=2\pi/\beta$ is the first positive Matsubara frequency.
 At a critical point, we expect 
$\xi_\tau(L) \propto L^z$.
Figure \ref{Fig:allcl}(e) plots $\xi_\tau$ for different system sizes at $s=1$ and $s=0.95$.
Fits yield $z=1.02(2)$ and $z=1.12(1)$, respectively.
These values of $z$ indicate
that, as $s \to 1$, Lorentz symmetry is gradually restored, signaling that the FP moves
towards and reaches $g = 0$ at $s = 1$. For a spin-$1/2$ system, a critical fixed point possessing both Lorentz and SU(2) symmetry is identified as the $\mathrm{SU}(2)_1$ CFT. Our data therefore provide strong evidence for our RG analysis at $s \approx 1$ regime.

\textit{Conclusion and discussion}--In this letter, we have studied the dissipative $J$-$Q_3$ chain using non-Abelian bosonization, RG analysis, and QMC simulations.  We have found continuous AFM-VBS transitions in both the sub-Ohmic and Ohmic regimes and propose an AFM-QLRO transition in the super-Ohmic regime. In the Ohmic case, the transition is governed by the $\mathrm{SU}(2)_1$ CFT, where spinons deconfine and an O(4) symmetry emerges. In the sub-Ohmic regime, the transition may also involve the spinon deconfinement at criticality, serving as a DQCP analogue without emergent $\mathrm{O}(4)$ symmetry.

However, unlike in the 2D VBS phase, where spinons are confined, spinons in the VBS phase of an isolated spin chain are deconfined.
Whether spinons remain deconfined in the dissipative VBS phase, rendering the AFM--VBS transition only an analogue of the 2D DQCP, or instead become confined, making the transition a genuine DQCP, remains an open question.
Further investigation from the perspective of spinons is therefore called for.

It is also worth noting that, for $s = 0.85$, QMC simulations yield $z \approx 2$\cite{supplemental}, rendering the perturbative RG analysis based on the $z = 1$ CFT unreliable. A more appropriate
approach would be to start from a free theory with $z = 2$ and carry out an RG analysis. Moreover, we find $C(L/2)$, $m_s^2$, $C_D(L/2)$, and $D^2$ at criticality are remarkably similar for different values of $s$, with $\Delta_D \approx 1/4$ throughout the sub-Ohmic regime. These observations may point to a fixed point in the strongly dissipative regime, characterized by
$\Delta_D \approx 1/4$, $\Delta_N \approx 1/2$ and $z \approx 2$. Further investigation along this
direction is also warranted.

\textit{acknowledgments}---
We acknowledge helpful discussions with Z.J. Wang and M. Weber.
L.L. and W.G. acknowledge support from the National Natural Science Foundation of China under Grant No.~12574252. S.C. acknowledges support from the Shandong Provincial Natural Science Foundation under Grant No.~ZR2024MA043. 
The authors also acknowledge the Supercomputing Center of Beijing Normal University.

\bibliographystyle{apsreve}
\bibliography{cs}

\end{document}


\title{Supplementary Materials for \\ `` Deconfined quantum critical point in a dissipative spin-1/2 chain'' }
\author{Longye Lu}
\affiliation{School of Physics and Astronomy, Beijing Normal University, and Key Laboratory of Multiscale Spin Physics (Beijing Normal University), Ministry of Education, Beijing 100875, China}
\author{Shifeng Cui}
\affiliation{Shandong Engineering Research Center of Aeronautical Materials and Devices, College of Aeronautical Engineering, Shandong University of Aeronautics, Binzhou 256603, China}
\author{Wenan Guo}
\email{waguo@bnu.edu.cn}
\affiliation{School of Physics and Astronomy, Beijing Normal University, and Key Laboratory of Multiscale Spin Physics (Beijing Normal University), Ministry of Education, Beijing 100875, China}

\maketitle

\setcounter{secnumdepth}{3}
\setcounter{section}{0}
\def\thesection{\Roman{section}}
\setcounter{figure}{0}
\renewcommand\thefigure{S\arabic{figure}}
\setcounter{equation}{0}
\renewcommand\theequation{S\arabic{equation}}
\setcounter{table}{0}
\renewcommand\thetable{S\Roman{table}}

In these supplemental materials, we present details of the non-Abelian bosonization of the dissipative $J$-$Q_3$ model, the derivation of the renormalization group (RG) equations, the analysis of RG flow,
the spin-wave analysis of the effect of $Q$ terms, along with the quantum Monte Carlo simulation technique details and additional numerical results.  
\section{Non-Abelian Bosonization and RG equations}
\subsection{Non-abelian Bosonization}
We consider the dissipative $J$-$Q_3$ model described by the total Hamiltonian $H=H_s+H_{sb}$, with
\begin{equation}
H_{\rm s}=-J\sum_{i=1}^N  C_{i, i+1}-Q \sum_i C_{i, i+1} C_{i+2, i+3} C_{i+4, i+5},
\end{equation}
and
\begin{equation}\label{h_sb}
   H_{\rm sb}=\sum_{i q} \omega_q \mathbf{a}_{i q}^{\dagger} \cdot \mathbf{a}_{i q}+\sum_{i q} g_q (\mathbf{a}_{i q}^{\dagger}+\mathbf{a}_{i q}) \cdot \mathbf{S}_i,
\end{equation}
where $\mathbf{a}_{iq}$ and $\mathbf{a}^\dagger_{iq}$ are three-component vectors of bosonic annihilation and creation operators, respectively, at site $i$ for a mode $q$ with frequency $\omega_q$, and $g_q$ is the spin-boson coupling. By integrating out the bosonic degrees of freedom exactly, we obtain the effective retarded interaction of the spin degrees of freedom:
\begin{equation}
    H_{\rm ret}=-\iint_0^\beta d\tau d\tau'\sum_{i}K(\tau-\tau')\mathbf{S}_i(\tau)\cdot \mathbf{S}_i(\tau'),
\end{equation}
where $K(\tau-\tau')$ is the propagator:
\begin{equation}
   K(\tau) =\frac{1}{\pi}\int_0^{\omega_c}d\omega J(\omega)\frac{e^{-\omega\tau}}{1-e^{-\beta \omega}}\sim \frac{\alpha}{|\tau|^{1+s}}.
\end{equation}
with $J(\omega)=2\pi\alpha \omega_c^{1-s}\omega^s$, where $s=1$ corresponds to the Ohmic bath, $s<1$ the sub-Ohmic bath, and $s>1$ the super-Ohmic bath. $\alpha$ is the dimensionless coupling constant, $\omega_c$ is the frequency cutoff, and $\beta$ is the inverse temperature. 

Now we bosonize the resulting Hamiltonian ${\cal H}=H_{\rm s}+H_{\rm ret}$. To preserve the $\mathrm{SU}(2)$ symmetry,  we adopt the non-Abelian bosonization method. 

The bosonization result of the Heisenberg term in $H_{\rm s}$, which is equivalent to an isolated Heisenberg chain, is well-known and yields precisely the $\mathrm{SU}(2)_1$ CFT plus a backscattering interaction term:
\begin{equation}
\label{hamiltonian}\begin{split}
    H_{\text{Heisenberg}}&=\frac{1}{6\pi}\int dxd\tau \left[\vec{J}_L(w)\cdot \vec{J}_L(w)+\vec{J}_R(\bar{w})\cdot \vec{J}_R(\bar{w})\right]\\
   &+\lambda_0 \int dxd\tau \vec{J}_L(w)\cdot\vec{J}_R(\bar{w}),
\end{split}
\end{equation}
where the first term is the Sugawara construction of the $\mathrm{SU}(2)_1$ energy-momentum tensor, and the second term describes a backscattering process, with $\lambda_0<0$ being the backscattering coupling constant. 
Here, the microscopic spin operator is expressed 
$\vec{S}(x,\tau) \sim \bigl( \vec{J}_L(x,\tau) + \vec{J}_R(x,\tau) \bigr) + (-1)^x \, \vec{n}(x,\tau)$, where $\vec{n}(x, \tau)$ is the N\'eel order parameter, $\vec{J}_{L, R}$ are the left and right chiral modes generating the $\mathrm{SU(2)}_{L,R}$ symmetries, respectively. 

The $Q$ term in $H_{\rm s}$ can be bosonized straightforwardly. 
Since the local valence bond solid (VBS) order parameter $v(x) \sim (-1)^j {\mathbf S}_j\cdot {\mathbf S}_{j+1}$, the 4-spin interaction in the $Q$-term can be written as $(v(x))^2\propto -a \vec{J}_L(w)\cdot\vec{J}_R(\bar{w})$.
Therefore, in fact, the role of the $Q$-term is to provide a positive modulation of the backscattering, leading $\lambda_0 \to \lambda$. 

Without dissipation, the RG treatment of the backscattering interaction $\lambda$ is well established in textbooks\cite{Giamarchi-1D}. With our sign convention, $\lambda<0$ is marginally irrelevant and flows to $\lambda=0$, which corresponds to the $\mathrm{SU}(2)_1$ CFT. A positive $\lambda$ is marginally relevant and flows under RG to a VBS phase, breaking lattice translational symmetry. This picture is consistent with numerical simulations of the $J$-$Q_3$ chain \cite{TangYing1dJQ3, Kedar1dJQ3}: for a small value of $Q$, the system remains in the critical phase (since $\lambda$ remains negative and flows to the $\mathrm{SU}(2)_1$ CFT); increasing $Q$ exceed a critical value $Q_{c,0}$, the system undergoes a phase transition into the VBS phase. The critical point $Q=Q_{c,0}$ corresponds to $\lambda=0$. 

Next, we proceed to bosonize $H_{\text{ret}}$. 
We have:
\begin{equation}
    \begin{split}
        \vec{S}(x,\tau) \cdot \vec{S}(x,\tau') \sim &
\; \vec{J}_L(x,\tau) \cdot \vec{J}_L(x,\tau') 
+ \vec{J}_L(x,\tau) \cdot \vec{J}_R(x,\tau') 
+ (-1)^x\, \vec{J}_L(x,\tau) \cdot \vec{n}(x,\tau') \\
&+ \vec{J}_R(x,\tau) \cdot \vec{J}_L(x,\tau') 
+ \vec{J}_R(x,\tau) \cdot \vec{J}_R(x,\tau') 
+ (-1)^x \, \vec{J}_R(x,\tau) \cdot \vec{n}(x,\tau') \\
&+ (-1)^x \, \vec{n}(x,\tau) \cdot \vec{J}_L(x,\tau') 
+ (-1)^x \, \vec{n}(x,\tau) \cdot \vec{J}_R(x,\tau') 
+ \, \vec{n}(x,\tau) \cdot \vec{n}(x,\tau'),
    \end{split}
\end{equation}
of which only the last term needs to be retained, for the following reasons.
\begin{enumerate}
    \item The oscillating terms containing the factor $(-1)^x$ (i.e., those involving  $J \cdot n$) vanish upon spatial integration.
    \item In $H_{\text{ret}}$,  the interaction involves an additional doubel time integral over a propagator $\alpha \int dx \, d\tau \, d\tau' \, |\tau|^{-(1+s)}$, whose scaling dimension is  $[\alpha]+(1+s)-1-2$. 
    Since an interaction of the form $J \cdot J$ has scaling dimension $2$, we find $[\alpha]=-[2-1-2+(1+s)] = -1+(1-s)$. For $s>0$, this is negative, rendering such terms irrelevant.
    \item For the last term, $n\cdot n$, recall that the scaling dimension of $\vec{n}$ at the $\mathrm{SU}(2)_1$ CFT is $1/2$. A dimensional analysis of $\alpha \int dx \, d\tau \, d\tau' \, \vec{n}(x,\tau)\cdot\vec{n}(x,\tau') / |\tau-\tau'|^{1+s}$ yields $[\alpha] = 1-s$. Thus, the coupling is relevant, marginal, and irrelevant for $s<1$, $s=1$, and $s>1$, respectively. 
\end{enumerate}

Finally, we obtain the non-Abelian bosonized form of our model 
\begin{equation}
\label{h_boso}
\begin{split}
    H&=\frac{1}{6\pi}\int dxd\tau \left[\vec{J}_L(w)\cdot \vec{J}_L(w)+\vec{J}_R(\bar{w})\cdot \vec{J}_R(\bar{w})\right]\\
   &+\lambda \int dxd\tau \vec{J}_L(w)\cdot\vec{J}_R(\bar{w}) -\tilde{\alpha}\int dx d\tau  d\tau'\frac{\vec{n}(x,\tau)\cdot\vec{n}(x,\tau') }{ |\tau-\tau'|^{1+s}},
\end{split}
\end{equation}
where the last term describes the retarded interactions.
This theory captures all the important interactions of our lattice model. 
The power-counting result obtained here is consistent with the statements in \cite{DissipationInducedOrder}, in support of the effective action we have derived.

\subsection{Conformal Perturbation Theory}
For the convenience of the calculations \cite{Cenke_xu_Neel_VBS}, we introduce a generalized free field $\vec{\phi}$ to reproduce the retarded interaction generated by dissipation, with
\begin{equation}
 H_\phi=\int d^2x d^2x'\phi^i(x,\tau)C_{ij}^{-1}(x,\tau;x',\tau')\phi^j(x',\tau'),
 \end{equation}
where
\begin{equation}\label{ret}
   C_{ij}(x,\tau;0,0)=\langle\phi^i(x,\tau)\phi^j(0,0)\rangle=\frac{\delta_{ij}\delta(x)}{|\tau|^{1+s}}.
\end{equation}
where $i,j=1,2,3$. 

The Hamiltonian of Eq.(\ref{h_boso}) is then written as 
\begin{equation}\label{h_total}
H=H_1+H_2+H_\phi,
\end{equation} 
with
\begin{align}\label{h1h2}
    H_1&=\frac{1}{6\pi}\int dxd\tau \left[\vec{J}_L(w)\cdot \vec{J}_L(w)+\vec{J}_R(\bar{w})\cdot \vec{J}_R(\bar{w})\right]
	+\lambda \int dxd\tau \vec{J}_L(w)\cdot \vec{J}_R(\bar{w}),\nonumber\\
	H_2&=g a^{\frac{s-1}{2}}\int dx d\tau \vec{\phi}(w,\bar{w})\cdot \vec{n}(w,\bar{w}).
\end{align}
Here, $H_2$ describes interactions between the N\'eel field and the
generalized free field $\vec{\phi}$. The factor $a^{(s-1)/2}$ renders the coupling $g$ dimensionless, where $a$ is the real space cutoff. 
In general, such a factor is $a^{\Delta_k-d}$ for a perturbation $g_k\int d^dx O_k$,  where $\Delta_k$ is the scaling dimension of the operator $O_k$.
A complex coordinate $w=\tau+ix$ is also introduced.
The model is symmetric under 
$g\rightarrow -g$, and  we restrict our discussion to the case  $g\geq0$ throughout the remainder of this paper.
After integrating out $\vec{\phi}$ in the partition function,  the resulting effective interaction is a retarded interaction of the form Eq. (\ref{h_boso}), with $\tilde{\alpha}\sim g^2$. Note that  a minus sign originates from the definition of the effective action, $e^{-S_{\text{eff}}}$, leading to the minus sign in Eq. (\ref{h_boso}).

Following standard treatment in conformal field theory, the one-loop beta function is \cite{Cardy_1996}\cite{Fradkin_2013}:
\begin{equation}
    \beta_k=\frac{dg_k}{dl}=(d-\Delta_k)g_k+ \sum_{ij}B_{ijk} g_ig_j
\end{equation}
with $B_{ijk}$ satisfying: 
\begin{equation}\label{bijk}
    B_{ijk} a^{\Delta_k-d}dl=-\frac{1}{2}C_{ijk} a^{\Delta_i+\Delta_j-2d}\times\int_{a<|x|<ae^{dl}} d^d x f(\vec{x}),
\end{equation}
where $g_1=g, g_2=\lambda$, $d$ is the spacetime dimension, $e^{dl}\approx 1+dl$ is scaling factor,
$\Delta_k$ is the scaling dimension of the operator $O_k$ in the perturbation $g_k\int d^2x O_k(x)$; 
$C_{ijk}$ is the \textbf{operator product expansion (OPE) coefficient} that can be obtained from the CFT of the starting fixed point; when $i\neq j$, a factor 2 should be multiplied to $C_{ijk}$ because of the symmetric contribution from $g_ig_j$ and $g_jg_i$ to the $\beta$-function; the shell integral over $f(\vec{x})$, performed in a disk $a<r<ae^{dl}$, comes from the singular factor generated by the OPEs, which will be explained later. It is worth pointing out that the integral on the right hand side of Eq. (\ref{bijk}) always contains a factor of $dl$, and it may yield an additional factor $a^X$, ultimately making the power of $a$ the same  on both sides of the equation.

To obtain the $\beta$-functions of model Eq.~(\ref{h_total}), we need the following OPEs: 
\begin{align}\label{ope1}
 \left(\sum_{a=1,2,3} J_L^a(\zeta) J_R^a(\bar{\zeta})\right)\left(\sum_{b=1,2,3} J_L^b(w) J_R^b(\bar{w})\right)&  \sim \frac{3}{4} \frac{1}{|\zeta-w|^4}-\frac{2}{|\zeta-w|^2} \sum_{a=1,2,3} J_L^a(w) J_R^a(\bar{w})\nonumber \\
& \quad+\frac{3}{2} \frac{1}{(\bar{\zeta}-\bar{w})^2} T_L(w)+\frac{3}{2} \frac{1}{(\zeta-w)^2} T_R(\bar{w})+\ldots,
\end{align}

\begin{align}\label{ope2}
\left[\phi^a(\zeta,\bar{\zeta})n^a(\zeta,\bar{\zeta})\right]\left[\phi^a(w,\bar{w})n^a(w,\bar{w})\right] &=\langle\phi^a(x,\tau)\phi^b(0,0)\rangle  n^a(\zeta,\bar{\zeta})n^a(w,\bar{w})\nonumber\\
&=\frac{\delta(x-x')}{|\tau-\tau'|^{1+s}}n^a(\zeta,\bar{\zeta})n^a(w,\bar{w})\nonumber\\
&=\frac{\delta(x-x')}{|\tau-\tau'|^{1+s}}\left[\frac{3}{2}\frac{1}{|\zeta-w|}\left(1+\frac{1}{2}(\zeta-w)^2 T_L(w)+\frac{1}{2}(\bar{\zeta}-\bar{w})^2 T_R(\bar{w})\right) \right.\nonumber\\
&\quad\left.+\frac{1}{2}|\zeta-w|J_L^a(w)J_R^a(\bar{w})\right],
\end{align}
and 
\begin{align}\label{ope3}
\left(\sum_{a=1,2,3} J_L^a(\zeta) J_R^a(\bar{\zeta})\right)\left(\sum_b n^b(w, \bar{w}) \Phi^b(w, \bar{w})\right) 
\sim \frac{1}{4} \frac{1}{|\zeta-w|^2}\left(\sum_b n^b(w, \bar{w}) \Phi^b(w, \bar{w})\right),
\end{align}
where $T_{L,R}\equiv \frac{1}{3}J^a_{L,R}J^a_{L,R}$ is the energy momentum tensor of $\mathrm{SU(2)}_1$ CFT. For small $\varepsilon=1-s\geq 0$, the OPEs truncated at this order provide an adequate approximation. Basically, these OPEs can be found in \cite{Cenke_xu_Neel_VBS}, with only a modification of the propagator $\langle\phi\phi\rangle$. 

The OPEs (\ref{ope1})-(\ref{ope3}) are sufficient to obtain the one-loop $\beta$-functions. For example, from Eq. (\ref{ope3}), we find that there is a $\lambda g$ contribution to $\beta_g=dg/dl$, with a coefficient
\begin{align}
    B_{\lambda g g} a^{(s-1)/2}dl&=-\frac{1}{2}\times 2 C_{\lambda g g}a^{(s-1)/2} \int_{a<|\zeta-w|<ae^{dl}} d^2 \zeta \frac{1}{|\zeta-w|^2}\nonumber\\
    &=-\frac{1}{2}\times 2\times a^{(s-1)/2}\frac{1}{4}\times \int_{a<r<ae^{dl}} d^2r \frac{1}{r^2}\nonumber\\ 
    &=-\frac{a^{(s-1)/2}}{4}\int_{a<r<ae^{dl}} dxd\tau\frac{1}{x^2+\tau^2}\nonumber\\
    &=-\frac{\pi}{2}a^{(s-1)/2} dl,
\end{align}
 which means $ B_{\lambda g g}=-\pi/2$. The other second order contributions to the $\beta$-function can be computed in the same way. 

Finally, we obtain the following  $\beta$-functions:
\begin{align}\label{beta_func}
    \beta_g=\frac{dg}{dl}&=\frac{1-s}{2}g-\frac{\pi}{2}\lambda g, \nonumber\\
    \beta_\lambda=\frac{d\lambda}{dl}&=2\pi\lambda^2-\frac{1}{2}g^2.
\end{align}
In the main text we introduced $\varepsilon=1-s$ to align with the language of $\varepsilon$-expansion. The leading-order terms can be obtained by dimensional analysis of Eq. (\ref{h_total}). 

It is clear that $g$ is marginal for the Ohmic case ($s=1$), while in the sub-Ohmic ($s<1$) and super-Ohmic ($s>1$) cases, $g$ is relevant and irrelevant, respectively. 
This finding matches the statement in \cite{DissipationInducedOrder}. 
\section{RG Flow Analysis}
In this section, we give a detailed analysis of the RG flows of Eqs. (\ref{beta_func}). 
\subsection{dynamical critical exponent}
We first determine the dynamical critical exponent \( z \) using the method introduced in \cite{antunes2026lifshitzcriticalpointsmeet}. 

Introducing the perturbation $S\rightarrow S+g^{\mu\nu}\int d^2x T_{\mu\nu}$, we obtain from the OPEs (\ref{ope1}) and (\ref{ope2}) the $\beta$-functions for the metric:
\begin{align}
    \beta^{ww}&=\frac{dg^{ww}}{dl}=-\frac{3}{4}g^2, \nonumber\\
    \beta^{\bar{w}\bar{w}}&=\frac{dg^{\bar{w}\bar{w}}}{dl}=-\frac{3}{4}g^2. 
\end{align}
At the new fixed points,  these $\beta$-functions can be non-zero(see later discussion of RG flow) and contribute to the trace of the stress-energy tensor via the following relation:
\begin{equation}
    T^{\mu}_{\mu}=\sum_{i}\beta_iO_i,
\end{equation}
where $O_i$ is the perturbing operator with $\beta_i$ being its $\beta$-function, which, in our model, becomes:
\begin{equation}
    T_{\tau\tau}+T_{xx}=-\frac{3}{4}g^2 T^{ww}-\frac{3}{4}g^2 T^{\bar{w}\bar{w}}.
\end{equation}
Note that all other $\beta_i$s vanish at a new fixed point. Transforming the right-hand side into Euclidean coordinates, we obtain the following equation for the stress-energy tensor:
\begin{equation}
    (1+\frac{3}{4}g^2)T_{\tau\tau}+T_{xx}=0.
\end{equation}
Comparing this to the Lifshitz-traceless condition  \cite{antunes2026lifshitzcriticalpointsmeet}, 
\[zT_{\tau \tau}+\delta^{i j} T_{i j}=0, \]
where $i$,$j$ are spatial indices, we obtain the dynamical exponent:
\begin{equation}\label{dynamical_exp_z}
    z=1+\frac{3}{4}g^2.
\end{equation}
Note that $g$ on the right-hand side of the equation should be evaluated at its fixed-point value $g^*$.  Eq. (\ref{dynamical_exp_z}) implies that a fixed point located at $g > 0$ always possesses a dynamic critical exponent $z > 1$. We also notice that the deviation of $z$ is 2nd order in perturbation, which is typical in a perturbative RG scheme\cite{antunes2026lifshitzcriticalpointsmeet}\cite{XuYCDQCnonlocal}.

\subsection{ N\'eel-VBS transition at sub-Ohmic dissipation ($s<1$) }

\begin{figure}[htbp]
    	\centering
	\includegraphics[width=0.44\textwidth]{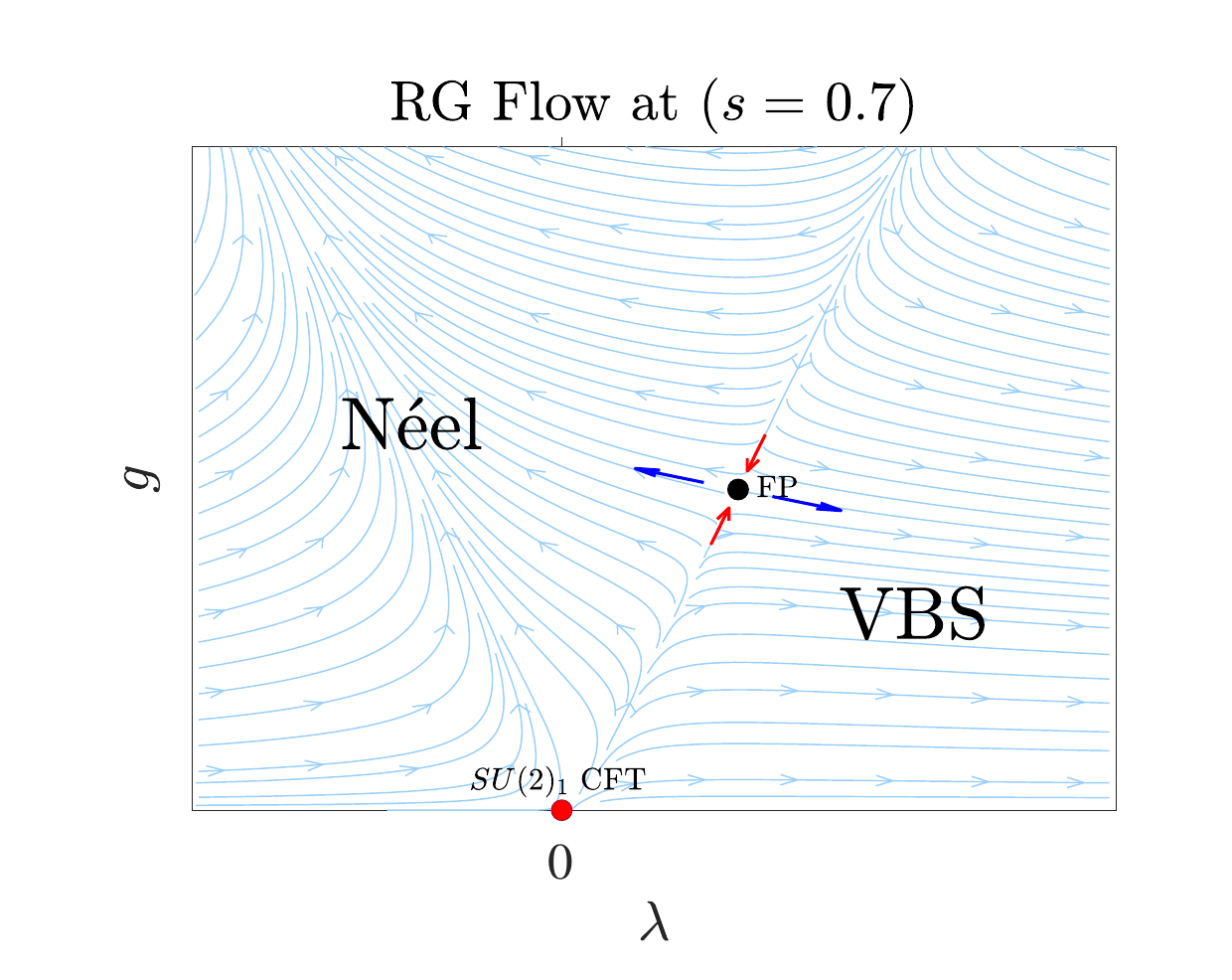}
	\caption{The RG phase diagram of Hamiltonian (\ref{h_total}) for sub-Ohmic ($s<1$) case, computed from the $\beta$-function (\ref{beta_func}). A new fixed point, labeled as FP, with dynamical exponent $z>1$ and distinct from the $\mathrm{SU}(2)_1$ CFT point, emerges on a critical surface that separates the N\'eel AFM phase and the VBS phase. The short arrows near FP are the eigen-directions.} 
	\label{Fig:rg_s=0.7}
\end{figure}

Based on the $\beta$-function Eq. (\ref{beta_func}), we obtain the phase diagram
for the sub-Ohmic regime $s < 1$, shown in  Fig. \ref{Fig:rg_s=0.7}.
The competition between the backscattering $\lambda$ and the dissipation $g$ leads to the emergence of a non-trivial fixed point (FP):
\begin{equation}
    \lambda^* = \frac{\varepsilon}{\pi}, \quad g^* = \frac{2\varepsilon}{\sqrt{\pi}},
\end{equation}
which sits on a critical surface separating the N\'eel and the VBS phases. 
Due to the fact that $g^* > 0$, this Fp is characterized by a dynamical critical exponent $z > 1$, according to Eq.(\ref{dynamical_exp_z}): 
\begin{equation}\label{exp_z}
    z=1+3(g^*)^2/4=1+3\varepsilon^2/\pi.
\end{equation} 

At this dissipative FP, the scaling dimensions of the N\'{e}el and VBS order parameters are modified by the finite couplings $(\lambda^*, g^*)$. 
Specifically, at $\mathrm{O}(\varepsilon)$, the scaling dimensions are given by:
\begin{equation}\label{eta}
    \Delta_N =\frac{1}{2}+\frac{\varepsilon}{2}, \quad \Delta_D = \frac{1}{2}-\frac{3\varepsilon}{2}.
\end{equation}
The derivation will be presented in Sec. \ref{sec:eta}.

Spin chains in the presence of Ohmic dissipation have
 been considered in the absence of the Berry phase \cite{Pankov_spinchain_RG, Sachdev_Nanowires} and verified by QMC simulations \cite{Werner_MC_spinchain_dissipation}. 
 Finite coupling to the bath is required to trigger a phase transition from a disordered phase at weak coupling to an ordered phase at strong coupling, with dynamical critical exponent $z=2-\eta$,
 which stems from the requirement of scale invariance at the fixed point. Applied to the model we consider, this constraint takes the form $2\Delta_N+(s-1)z=1$. To leading order in $\varepsilon$, with $z=1+\mathrm{O}(\varepsilon^2)$, this constraint yields $\Delta_N=\frac{1}{2}+\frac{\varepsilon}{2}$, which agrees with Eq. (\ref{eta}) obtained from the conformal perturbation calculation. This demonstrates that the constraint relation is automatically preserved at every order of our RG calculation. 

These results indicate that, compared to the $\mathrm{SU}(2)_1$ CFT, the correlation of N\'eel order at the new fixed point is suppressed with $\Delta_N>1/2$, while the VBS correlation is enhanced with $\Delta_D<1/2$, similar to the model studied in \cite{Cenke_xu_Neel_VBS}. 

Linearizing Eq. (\ref{beta_func}) at $(\lambda^*,g^*)$, we have the correlation length exponent
\begin{equation}\label{nu}
    \nu= 1/[(2+\sqrt{6})(1-s)].
\end{equation}

 It is worth noting that the phase diagram obtained here is similar to the one found in Ref.\cite{Cenke_xu_Neel_VBS} with $g_v=0$, and our $(1-s)/2$ plays the role of their $\varepsilon_n$,
 where the system has nonlocal interactions due to coupling to bulk critical fluctuations.

\subsection{Merging of the N\'eel-VBS fixed point into the $\mathrm{SU}(2)_1$ CFT at the Ohmic dissipation ($s=1$) }
\begin{figure}[h]
    	\centering
	\includegraphics[width=0.44\textwidth]{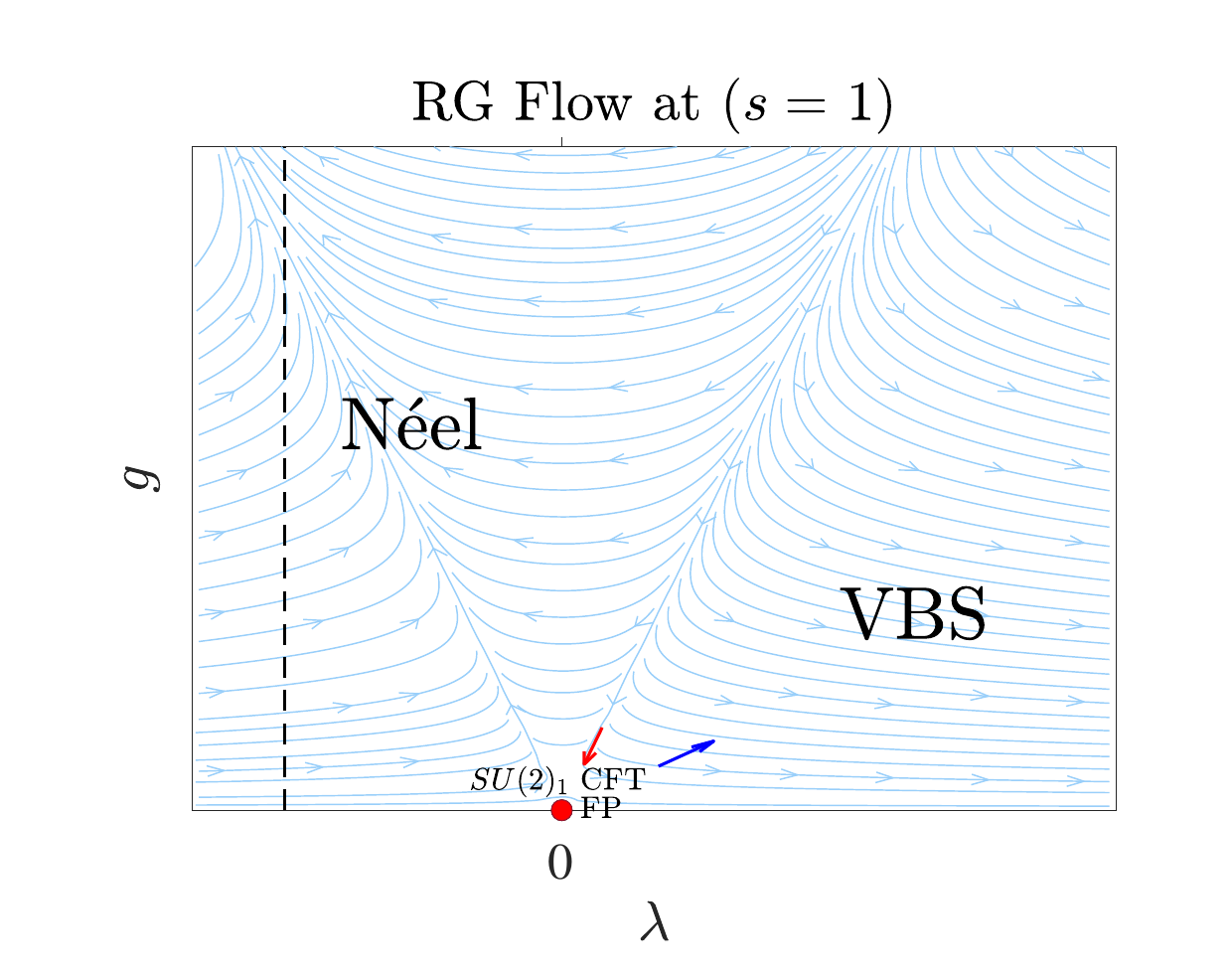}
	\caption{The RG phase diagram of Hamiltonian (\ref{h_total}) for Ohmic dissipation ($s=1$), computed from the $\beta$-function (\ref{beta_func}). The short arrows near the fixed point(FP) are the eigen-directions. The black dashed line indicates the RG starting points of a dissipative Heisenberg chain $\lambda_0<0$. 
    FP and $\mathrm{SU(2)}_1$ fixed points merge into a single fixed point.
    } 
	\label{Fig:rg_s=1}
\end{figure}

For the Ohmic dissipation, we obtain the phase diagram illustrated in Fig. \ref{Fig:rg_s=1}. 
Compared to the sub-Ohmic case, the N\'eel-VBS critical line still exists; 
however, the fixed point FP and the $\mathrm{SU}(2)_1$ fixed points merge into a single fixed point, which leads the N\'eel-VBS transition to be a deconfined quantum phase transition controlled by $\mathrm{SU}(2)_1$ CFT.

Right on the critical line, the RG flow is marginally irrelevant toward the origin, which explains the multiplicative logarithmic correction to the power law decay of the N\'eel order correlation found in our numerical simulations.
We will present details on the multiplicative logarithmic correction in Sec. \ref{sec:logcor}.

At $s=1$, the exponents $\Delta_N$ and $\Delta_D$ given by Eq. (\ref{eta}) both tend to $1/2$, and with $g^*\sim (1-s)=0$, Eq.(\ref{exp_z}) implies $z=1$, indicating the emergence of Lorentz symmetry. This is consistent with a flow back to the $\mathrm{SU}(2)_1$ fixed point. 

By including dissipation, this fixed point describes a quantum phase transition, instead of a critical phase, which is an ideal realization of the N\'eel-VBS deconfined quantum critical point: it hosts deconfined spinons and an emergent $O(4)$ symmetry.
Our RG analysis not only yields the RG phase diagram that matches the QMC simulations, but also explains  
the numerical results $\Delta_N=0.50(1)$, $\Delta_D=0.50(1)$, $z=1.02(2)$ at the N\'eel-VBS critical point for the Ohmic dissipation.

\subsection{Crossover behavior and $Q$ term enhancement  of AFM order}
The RG flow can also explain the crossover phenomena in \cite{DissipationInducedOrder}, and the peculiar 'Q-term strengthens N\'eel order' phenomena we observe in the QMC simulation. 
They can be understood as follows: 
The isolated Heisenberg model is described by the $\mathrm{SU(2)}_1$ CFT and an extra backscattering process with $\lambda_0<0$, which is marginally irrelevant and gives rise to logarithmic correction to the power-law decay of the spin correlation function. With dissipation turned on, the system lies somewhere along the black dashed line in Fig. \ref{Fig:rg_s=1}. From the flow diagram, we observe a distinct bending and reversal of the RG trajectories. The RG time at which the bending point is approached defines a crossover scale $L_c$; for system sizes smaller than $L_c$, the system behaves like a critical Heisenberg chain, while beyond $L_c$, LRO develops. 
As the dissipation strength increases, the system moves closer to the main trajectory, and the 
bending and reversal eventually disappear. Therefore, $L_c$ depends on $\alpha$.

Such behavior has been reported in the study of the Ohmic dissipative Heisenberg chain \cite{DissipationInducedOrder}, where it was attributed to the marginal relevance of Ohmic dissipation. We show here---supported by QMC simulation results presented in Fig. \ref{Fig:R_subo}
--that this crossover behavior persists in the sub-Ohmic case, where dissipation
is relevant at the limit $g=0$.

 Moreover, we believe that the enhancement of the N\'{e}el order by the $Q$ term, observed in the strong-dissipation regime, is also connected to this.
Increasing $Q$ moves the system along the positive $x$-direction, as shown in Fig.~\ref{Fig:rg_s=1}, thereby pushing it closer to the main trajectory that converges to the N\'eel AFM fixed point. This explains the observed $Q$-induced enhancement of N\'eel ordering in our QMC simulations, which in turn corroborates the RG flow analysis. 
This phenomenon can also be explained by our spin-wave analysis, to be presented in Sec. \ref{sec:SW}. Importantly, such enhancement only occurs in the presence of non-vanishing dissipation.

\subsection{ Analytical derivation of  the scaling dimentsions $\Delta_N$ and $\Delta_D$}
\label{sec:eta}
The scaling dimension $\Delta$ of an operator $\Phi $ can also be computed from conformal perturbation theory. Suppose $S_0$ is the unperturbed CFT, which is our starting fixed point, and operator $\Phi$ is a scaling operator with scaling dimension $\Delta_0$ in this CFT, i.e.:
\begin{equation}
    \langle \Phi(r) \Phi(0)\rangle_0=\frac{1}{r^{2\Delta_0}}
\end{equation}
The subscript $0$ denotes the expectation value computed with respect to the action $S_0$, e.g., for the N\'eel and VBS order parameter, $\Delta_{N,D;0}=1/2$. Now we add a perturbation, $S_1=g_O\int d^2x O(x)$ with $O(x)$ has scaling dimension $\Delta$. the correlation function becomes:
\begin{align}\label{nnlambda}
    \langle \Phi(r) \Phi(0)\rangle_\lambda&=\frac{\int D\phi e^{-S_0[\phi]-S_1}\Phi(r) \Phi(0)}{\int D\phi e^{-S_0[\phi]-S_1}}\nonumber\\
 &\approx \langle \Phi(r) \Phi(0)\rangle_0-g_O I
\end{align}
with 
\[
I=\int d^2x \langle O(x)\Phi(r) \Phi(0)\rangle_0.
\]
The integral $I$ in the 1st order correction can now be computed using the OPE, where $O(x)$ can contract with either $\Phi(r)$ or $\Phi(0)$, depending on which one is closer in distance:
\begin{align}
    I &=\int_{a<|x|<r} d^2x \frac{C_{O\Phi\Phi}}{|x|^2}\langle \Phi(r)\Phi(0)\rangle_0+\int_{a<|x
    -r|<r} d^2x \frac{C_{O\Phi\Phi}}{|x-r|^2}\langle \Phi(r)\Phi(0)\rangle_0. 
\end{align}
Note that in our case the singular part of the OPE has power $2$, which is the only situation that needs to be considered in the calculation for this model. The two integrals mentioned above are identical and evaluate to:
\begin{equation}
    \int_a^r \frac{C_{O\Phi\Phi}}{\rho^2} 2\pi \rho d\rho=2\pi C_{O\Phi\Phi} \ln{\frac{r}{a}},
\end{equation}
which yields $I=4\pi C_{Onn} \ln{\frac{r}{a}}\langle  \Phi(r) \Phi(0)\rangle_0$. Plugging it into Eq. (\ref{nnlambda}) yields
\begin{equation}\label{nn1}
    \langle \Phi(r) \Phi(0)\rangle_{g_O}\approx \langle \Phi(r) \Phi(0)\rangle_0\left[1-4\pi g_O C_{O\Phi\Phi} \ln{\frac{r}{a}}\right].
\end{equation}
Writing $\Delta=\Delta_0+\delta\Delta$, we have, to the first order,
\begin{equation}\label{nn2}
  \langle\Phi(r) \Phi(0)\rangle_{g_o}=\frac{1}{r^{2\Delta_0}}\frac{1}{r^{2\delta\Delta}}=\frac{1}{r^{2\Delta_0}}(1-2\delta\Delta\ln r).
\end{equation}
Comparing Eqs. (\ref{nn1}) and (\ref{nn2}), we find
\begin{equation}\label{delta_n}
    \delta \Delta=2\pi g_O C_{O\Phi\Phi}.
\end{equation}

Now we go back to our model (\ref{h_total}). The perturbations we consider are $g$ and $\lambda$. For the N\'eel order parameter $\Phi=\vec{n}$, since $C_{gnn}=0$, we only need to consider $\lambda$. From Eq.~(\ref{ope3}) we know $C_{\lambda n n}=1/4$; therefore,
\begin{equation}
    \delta\Delta_N=2\pi\lambda^*/4=\frac{1-s}{2},
\end{equation}
which yields $\Delta_N=(1/2+(1-s)/2)=1/2+\varepsilon/2$.

For the correction to $\Delta_v$, we need the following OPE \cite{Cenke_xu_Neel_VBS}:
\begin{align}\label{ope4}
& \left(\sum_{a=1,2,3} J_L^a(z) J_R^a(\bar{z})\right) v(w, \bar{w})\nonumber \\
& \sim -\frac{3}{4} \frac{1}{|z-w|^2} v(w, \bar{w}) ,
\end{align}
which yields $C_{\lambda v v}=-3/4$. Note that $v(w,\bar{w})$ denotes the VBS order parameter. Thus we have
\begin{equation}
    \delta\Delta_D=-3\pi\lambda^*/2=\frac{-3(1-s)}{2},
\end{equation}
which leads to $\Delta_D=1/2-3(1-s)/2=1/2-3\varepsilon/2$.

\subsection{Multiplicative logarithmic correction to scaling for Ohmic dissipation}
\label{sec:logcor}
The logarithmic correction can be derived from the Callan-Symanzik equation\cite{Victor_logarithmic_1999, I_Affleck_1989}. We recall the Callan-Symanzik equation\cite{2021qftc_book_Zinn-Justin}:
\begin{equation}\label{cs_eq}
    \left[\frac{\partial}{\partial l}+\beta(\lambda)\frac{\partial}{\partial\lambda}+2\gamma_n(\lambda)\right]G(r,\lambda)=0.
\end{equation}
Here $\gamma_n=\delta\Delta$ in eq(\ref{delta_n}). Since we know that $\langle n(r) n(0)\rangle_0=1/r$, the solution of Eq.~(\ref{cs_eq}) can be written as:
\begin{equation}\label{solu_cs}
    G(r)\sim \frac{1}{r}\exp\left(-2\int_{\lambda_0}^{\lambda(r)}\frac{\gamma_n(\lambda)}{\beta(\lambda)}d\lambda\right).
\end{equation}
For the case $s=1$, suppose we move along the critical line in Fig.\ref{Fig:rg_s=1}, along which $g^2=5\pi \lambda^2$. We obtain $\beta_\lambda=-\pi\lambda^2/2$. Then 
\begin{align}
-2\int_{\lambda_0}^{\lambda(r)}\frac{\gamma_n(\lambda)}{\beta(\lambda)}d\lambda&=2\int_{\lambda_0}^{\lambda(r)} \frac{d\lambda}{\lambda}
    =2\ln \frac{\lambda(r)}{\lambda_0},
\end{align}
which leads to 
\begin{equation}\label{solu2_cs}
    G(r)\sim \frac{1}{r}\exp\left(2\ln\frac{\lambda(r)}{\lambda_0}\right)=\frac{1}{r}\left(\frac{\lambda(r)}{\lambda_0}\right)^2.
\end{equation}
Here, $\lambda(r)$ can be computed from the beta function:
\begin{equation}\label{beta_lam_critical}
    \frac{d\lambda}{dl}=-\frac{\pi}{2}\lambda^2.
\end{equation}
Therefore, we have 
\begin{equation}
    \int_{\lambda_0}^{\lambda(r)}\frac{d\lambda}{\lambda^2}=-\frac{\pi}{2}\int_{\ln a}^{\ln r} dl,
\end{equation}
which leads to
\begin{equation}
    \frac{\lambda(r)}{\lambda_0}=\frac{1}{1+\lambda_0\frac{\pi}{2}\ln\frac{r}{a}}.
\end{equation}
Plugging this into Eq.~(\ref{solu2_cs}), we have:
\begin{equation}
  G(r)\sim\frac{\left(\ln r\right)^{-2}}{r}.
\end{equation}
Our QMC results reveal the presence of a multiplicative logarithmic correction with an exponent $-0.5$ to the power decay of the correlation function. While the trend of suppressing N\'eel order correlation agrees with the theory, however, the exponent differs from our theoretical value of $-2$.
The reasons for this difference could be  some unknown operator mixing or higher order effects. 
For the correlation of VBS order parameter $\langle v(r)v(0)\rangle$, following same calculation we have a $(\ln r)^{-6}$, whose exponent is 3 times the logarithmic correction of $G(r)$'s. This is because in $\mathrm{SU(2)_1}$ algebra, $C_{\lambda vv}=-3C_{\lambda nn}$.

\section{Spin-Wave Analysis of the effect of $Q$ terms}
\label{sec:SW}
In this section, we present spin-wave theroy analyssi of the effect of $Q$ terms in our dissipative $J$-$Q_3$ model. 
We begin with a brief review of the spin-wave analysis in \cite{DissipationInducedOrder}. 
Using the Holstein-Primakoff transformation for the A and B sublattices and keeping terms  
up to $\mathcal{O}(S^0)$, the dissipative AFM Heisenberg model can be written as
\begin{align}\label{hsb_sw}
\hat{H}_{\mathrm{s}}&=\frac{L q J S^2}{2}+J S \sum_{\langle i, j\rangle}\left(\hat{A}_i \hat{B}_j+\hat{A}_i^{\dagger} \hat{B}_j^{\dagger}\right)+q J S \sum_{i \in \mathrm{~A}} \hat{A}_i^{\dagger} \hat{A}_i\nonumber\\&+q J S \sum_{j \in \mathrm{~B}} \hat{B}_j^{\dagger} \hat{B}_j+\mathcal{O}\left(S^0\right),
\end{align}
where the retarded interaction term, which arises from integrating out the bosonic degrees of freedom, is not shown.

To see if the spin-wave correction destroys the long-range AFM order, we only need to calculate the expectation value of boson occupation number:
\begin{equation}\label{ns}
\begin{split}
N_S&=\frac{1}{L \beta} \int_0^\beta d \tau\left[\sum_{i \in \mathrm{A}}\left\langle\hat{A}_i^{\dagger}(\tau) \hat{A}_i(\tau)\right\rangle
+\sum_{j \in \mathrm{B}}\left\langle\hat{B}_j^{\dagger}(\tau) \hat{B}_j(\tau)\right\rangle\right]\\
&=\frac{2}{\pi} \int_0^{\omega_{\mathrm{c}}} d \omega \frac{2 J S+\tilde{\alpha} \omega^s}{\sqrt{\left(\omega^2+4 J S \tilde{\alpha} \omega^s+\tilde{\alpha}^2 \omega^{2 s}\right) \left[(2 J S)^2+\omega^2+4 J S \tilde{\alpha} \omega^s+\tilde{\alpha}^2 \omega^{2 s}\right]}}.
\end{split}
\end{equation}

Now we consider the effect of $Q$-term. For simplicity, we consider the $Q_2$ interaction:
\begin{align}
    H_Q&=-Q\sum_{i}(\textbf{S}_i\cdot \textbf{S}_{i+1})(\textbf{S}_{i+2}\cdot \textbf{S}_{i+3})\nonumber\\
    &=-Q\sum_{i\in A}(\textbf{S}_i^A\cdot \textbf{S}_{i+1}^B)(\textbf{S}_{i+2}^A\cdot \textbf{S}_{i+3}^B)-Q\sum_{i\in B}(\textbf{S}_i^B\cdot \textbf{S}_{i+1}^A)(\textbf{S}_{i+2}^B\cdot \textbf{S}_{i+3}^A)
\end{align}
For the summation over sublattice A, we have:
\begin{align}\label{Asummation}
& -Q\sum_{i\in A}(\textbf{S}_i^A\cdot \textbf{S}_{i+1}^B) (\textbf{S}_{i+2}^A\cdot \textbf{S}_{i+3}^B)\nonumber \\
&=-Q\sum_{i\in A}
  \left\{
    \left[
      \frac{1}{2}\left(S_{i}^{A+}S_{i+1}^{B-}+S_{i}^{A-}S_{i+1}^{B+}\right)
      +S_{i}^{A,z}S_{i+1}^{B,z}
    \right]
    \left[
      \frac{1}{2}\left(S_{i+2}^{A+}S_{i+3}^{B-}+S_{i+2}^{A-}S_{i+3}^{B+}\right)
      +S_{i+2}^{A,z}S_{i+3}^{B,z}
    \right]
  \right\}\nonumber\\
&=-Q\sum_{i\in A}
  \left\{
    \left[
      S\left(A_{i} B_{i+1}+A_{i}^\dagger B_{i+1}^\dagger\right)
      +(S-n_{i}^A)(-S+n_{i+1}^B)
    \right]
    \left[
      S\left(A_{i+2} B_{i+3}+A_{i+2}^\dagger B_{i+3}^\dagger\right)
      +(S-n_{i+2}^A)(-S+n_{i+3}^B)
    \right]
  \right\}\nonumber\\
&=-Q\sum_{i\in A}
  \left[   S^4
    +S^3\left(
      A_{i} B_{i+1}+A_{i}^\dagger B_{i+1}^\dagger
      +A_{i+2} B_{i+3}
     +A_{i+2}^\dagger B_{i+3}^\dagger +n_{i}^A+n_{i+1}^B+n_{i+2}^A+n_{i+3}^B
    \right)
    +\mathcal{O}\left(S^2\right),
  \right]
\end{align}
where $n_i^{A(B)}$ is the boson number operator $A_i^\dagger A_i$ ($B_i^\dagger B_i$). 
Besides the constant term, the leading correction from $Q$ term is proportional to $S^3$, which is beyond the leading order $S^1$ in Eq.(\ref{hsb_sw}). 
We therefore set $Q\sim 1/S^2$, so that these corrections are of order $S^2$ and $S^1$. 
Absorbing the $\mathcal{O}(S^3)$ correction in Eq.~(\ref{Asummation}) into the $O(S^1)$ terms of Eq.~(\ref{hsb_sw}) shifts the coefficient: $JS\rightarrow  (JS+QS^3)= JS(1+QS^2/J)\equiv J\tilde{S}$. 
Repeating the calculation of $N_s$ with the effect of $Q$-term interaction expressed as $JS\rightarrow J\tilde{S}$ in Eq.(\ref{ns}), we obtain the bosonic occupation number to the leading order in $\omega$:
\begin{equation}
    N_s=\fc{2}{\pi}\int_0^{\omega_{c}}d\omega\fc{1}{\sqrt{\omega^2+4J\tilde{S}\tilde{\alpha}\omega^s}},
\end{equation}
where the low-frequency contribution is:
\begin{equation}
    \frac{1}{\pi}\frac{1}{\sqrt{J\tilde{S}\tilde{\alpha}}}\frac{1}{\sqrt{\omega^s}}.
\end{equation}
The fact that $\tilde{S}=S(1+QS^3/J)>S$ reduces the low frequency contribution to $N_s$ compared to $Q=0$, leading to the enhancement of long-range AFM order in presence of $Q$-term interactions.

From Eq.(\ref{ns}), we also note that at $\alpha=0$ this correction cancels out and does not enhance the AFM order. Finite dissipation is required for the leading corrections from $Q$-term interactions to suppress the fluctuation $N_S$, thereby enhancing the long-range AFM order.

\section{Quantum Monte Carlo simulations}
\subsection{Convergence to ground state properties in QMC simulations}

\begin{figure}[htbp]
\includegraphics[width=0.7\textwidth]{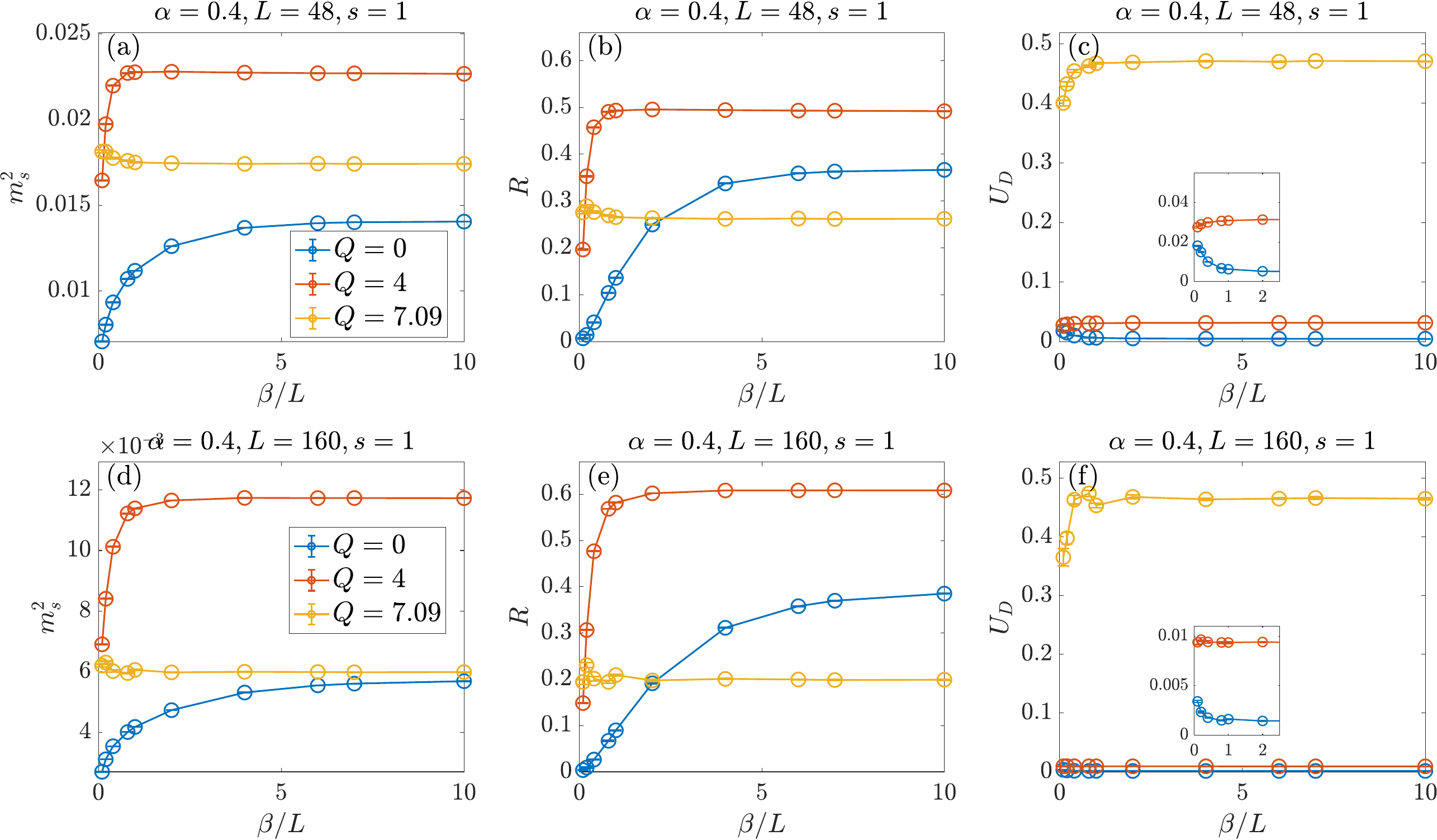}
\caption{The convergence analysis of ground-state properties for different $Q$ at $\alpha=0.4, L=48, 160$, and $s=1$. The required convergence ratio, $\beta/L$, is approximately the same for all system sizes in the range $L=24$ to $L=160$, and results for other system sizes are not shown.}
\label{beta_conver_ohmic}
\end{figure}

To study the ground-state properties of our model, we first make sure the observables are converged 
in the zero-temperature limit for each parameter set.  
Figure \ref{beta_conver_ohmic} shows the inverse-temperature dependence of the squared
AFM order parameter $m_s^2$, correlation ratio $R$, and VBS Binder cumulants $U_D$ for different value of $Q$s at $\alpha=0.4$ and $L=48, 160$, 
for Ohmic dissipation $s=1$.
In the vicinity of AFM-VBS critical point $Q_c\approx 7.09$, $\beta/L=6$ is sufficient to achieve convergence for all system sizes.

For the sub-Ohmic case, the dynamic exponent $z>1$ is expected at the VBS-AFM transition point. To reach the ground-state, one may set $\beta \sim L^z$. However, the value of $z$ is unknown a priori.
To obtain the correct temperature scaling (i.e., the proper choice of $\beta$), one should examine the temperature convergence case by case and incorporate the fit of $\xi_\tau \sim L^z$ to the data as a reference.

\begin{figure}[htbp]
\includegraphics[width=0.7\textwidth]{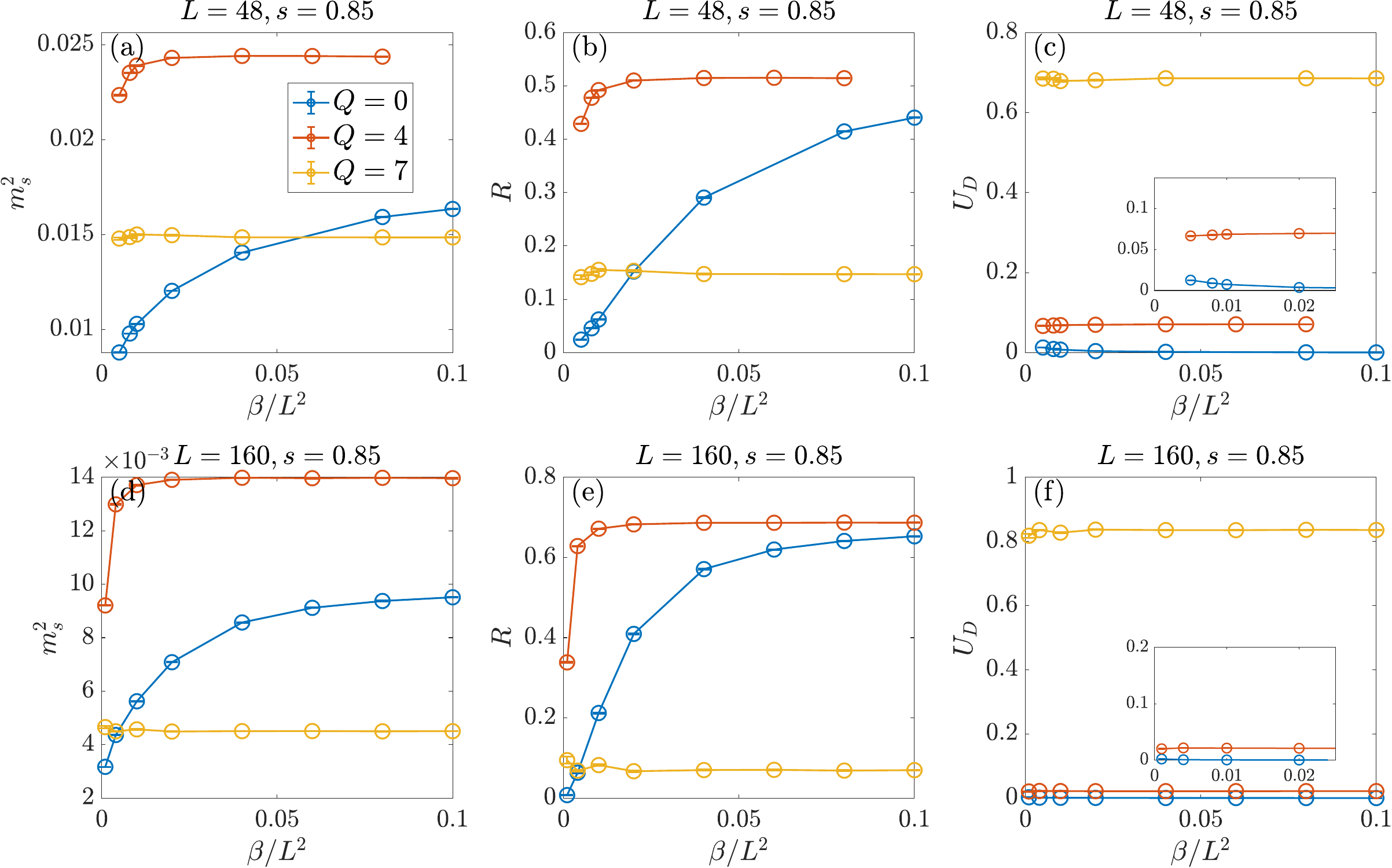}
\caption{
Temperature convergence of observables for $s=0.85$, $\alpha=0.4$, and various $Q$.
(a)--(c) show $m_s^2$, $R$, and $U_D$ for $L=48$;
(d)--(f) show the same quantities for $L=160$.
}
\label{beta_conver_s0.85}
\end{figure}

\begin{figure}[htbp]
\includegraphics[width=0.7\textwidth]{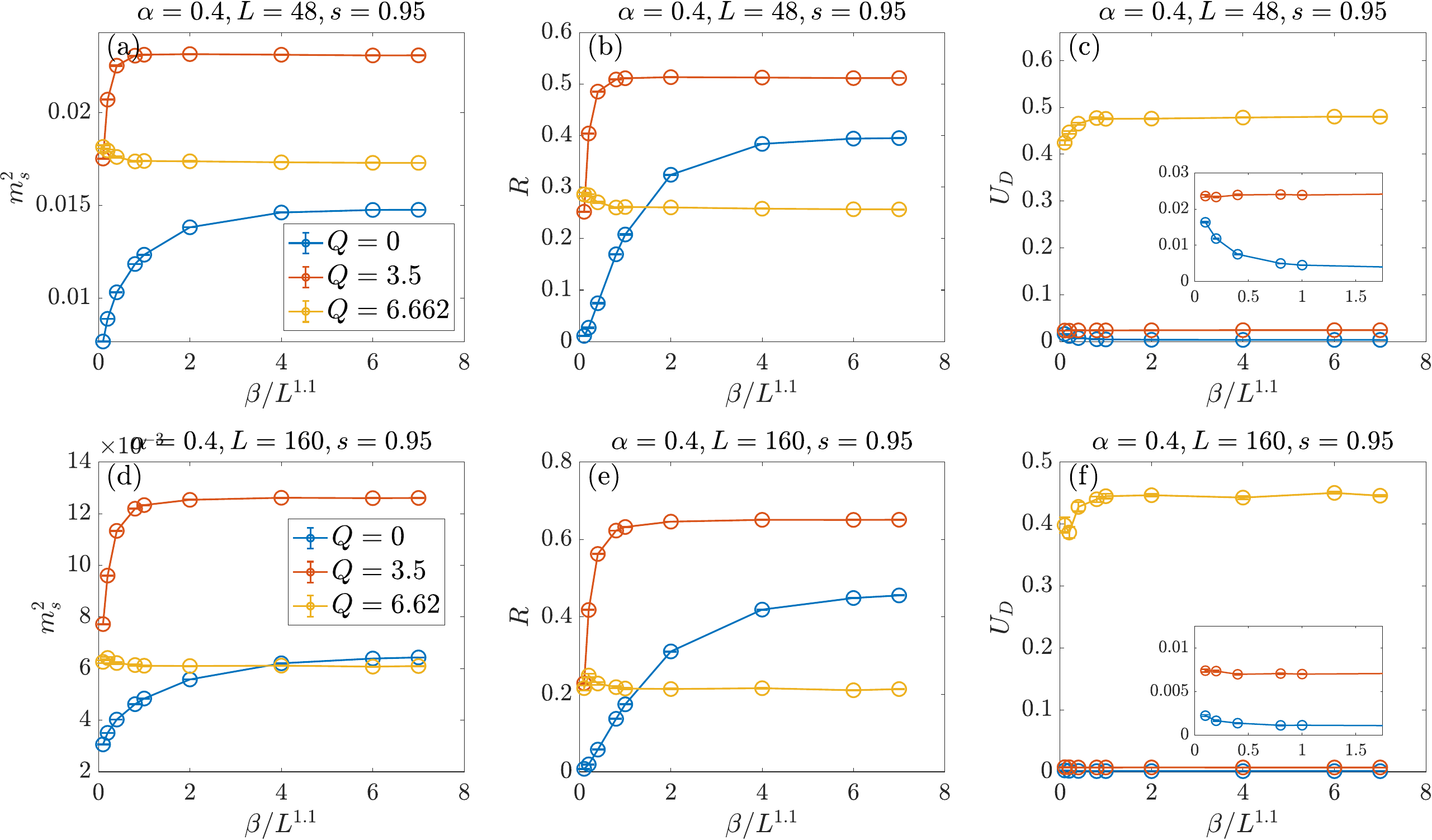}
\caption{
Temperature convergence of observables for $s=0.95$, $\alpha=0.4$, and various $Q$.
(a)--(c) show $m_s^2$, $R$, and $U_D$ for $L=48$;
(d)--(f) show the same quantities for $L=160$.
In this case $\beta=5L^{1.1}$ is sufficient for convergence to the ground state.}
\label{beta_conver_s0.95}
\end{figure}

Figures \ref{beta_conver_s0.85} and \ref{beta_conver_s0.95} show the inverse-temperature dependence of several observables for various $Q$ at two representative system sizes, for $s=0.85$ and $s=0.95$, respectively.
We find that, near $Q_c$, $\beta/L^{1.1}=5$ for $s=0.95$ and $\beta/L^{2}=0.05$ for $s=0.85$ are sufficient for convergence to the ground state. The temperature convergence behavior varies little across different physical observables and system sizes $L$. We further observe that, in all cases considered, convergence in $Q>0$ region is consistently faster than that in the $Q=0$ region. This indicates that the dynamical exponent $z$ at $Q_c$ is generally smaller than that at $Q=0$, consistent with the CFT prediction. For the Ohmic case, this may be interpreted as an indication that Lorentz symmetry is gradually being restored.

\begin{figure}[h]
\includegraphics[width=0.7\textwidth]{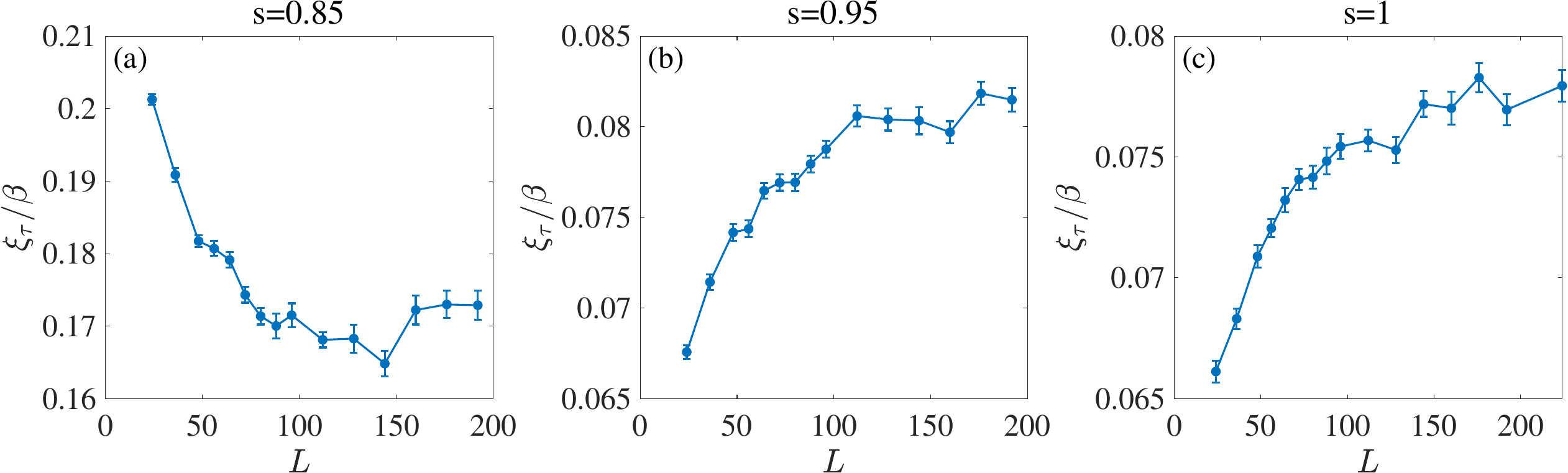}
\caption{$\xi_\tau/\beta$ at the AFM-VBS critical points with $\alpha=0.4$.(a) Sub-Ohmic case $s=0.85$ at $Q_c=5.962$. (b) Sub-Ohmic case $s=0.95$ at $Q_c=6.665$. (c) Ohmic case $s=1$ at $Q_C=7.095$}
\label{xitaubeta}
\end{figure}

As another way to check whether the chosen temperature is low enough to reach the ground state, it is crucial to ensure $\xi_\tau/\beta < 0.5$. Figures \ref{xitaubeta}(a)-(c) show the size-dependent behavior of $\xi_\tau/\beta$ for the Ohmic and sub-Ohmic cases at the AFM-VBS critical points, respectively.   
We have set $\beta=0.05L^2$ for $s=0.85$, $\beta=5L^{1.1}$ for $s=0.95$, and $\beta=6L$ for the Ohmic bath.
In all cases, we see $\xi_\tau/\beta$ stays smaller than 0.5 as the system size $L$ becomes large. Moreover, in all three cases, as the system size increases, $\xi_\tau/\beta$ settles onto a stable plateau, indicating that $\xi_\tau$ and $\beta$ vary synchronously with system sizes, which demonstrates the correctness of our choice of the inverse temperature $\beta$.

\subsection{AFM-VBS transition in Sub-Ohmic($s=0.85$) dissipation}
\label{subohmicQMC}
In this subsection, we present large-scale simulation results in the sub-Ohmic region.

\begin{figure}[htbp]
    \centering
     \includegraphics[width=0.30\textwidth]{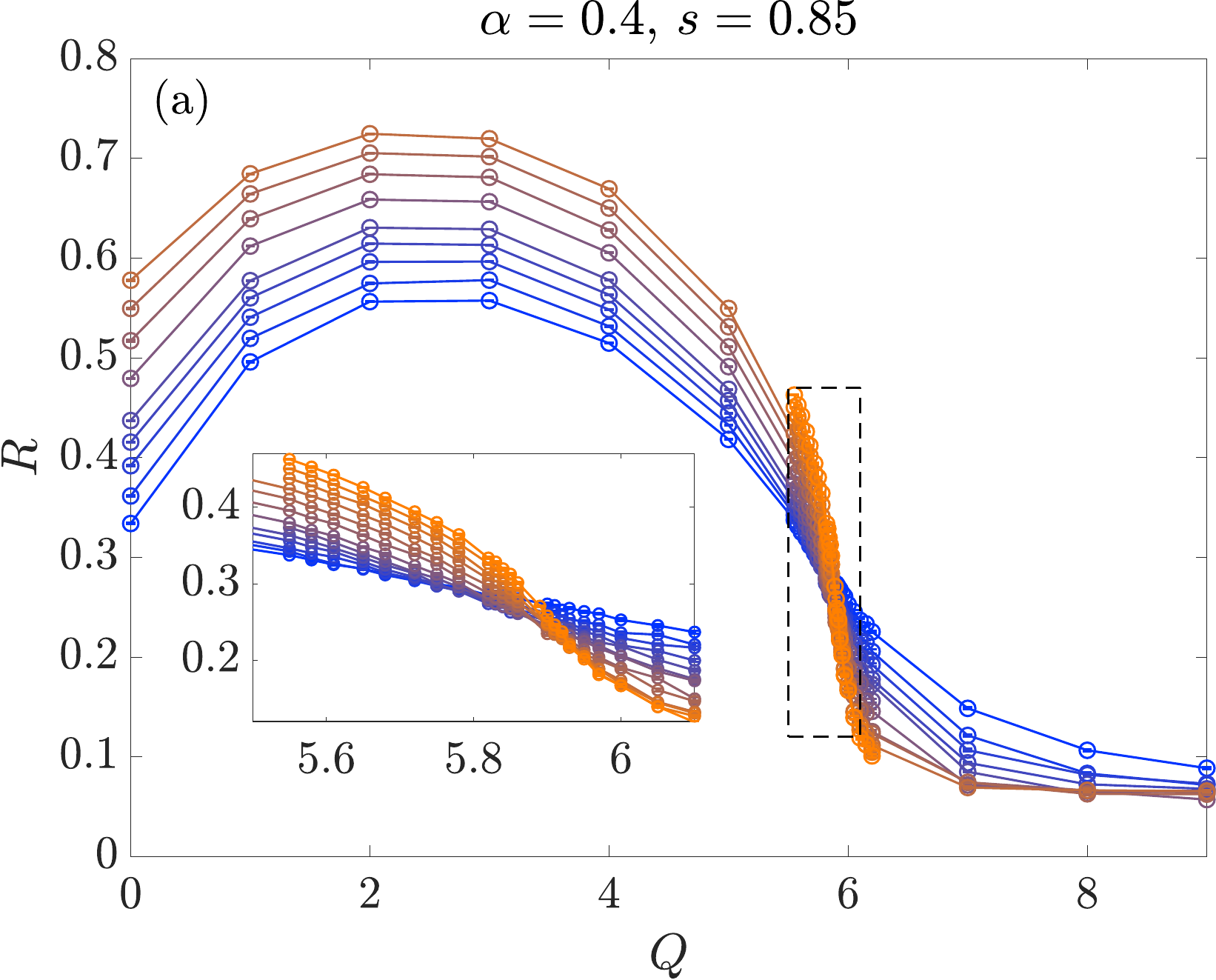}
    \includegraphics[width=0.30\textwidth]{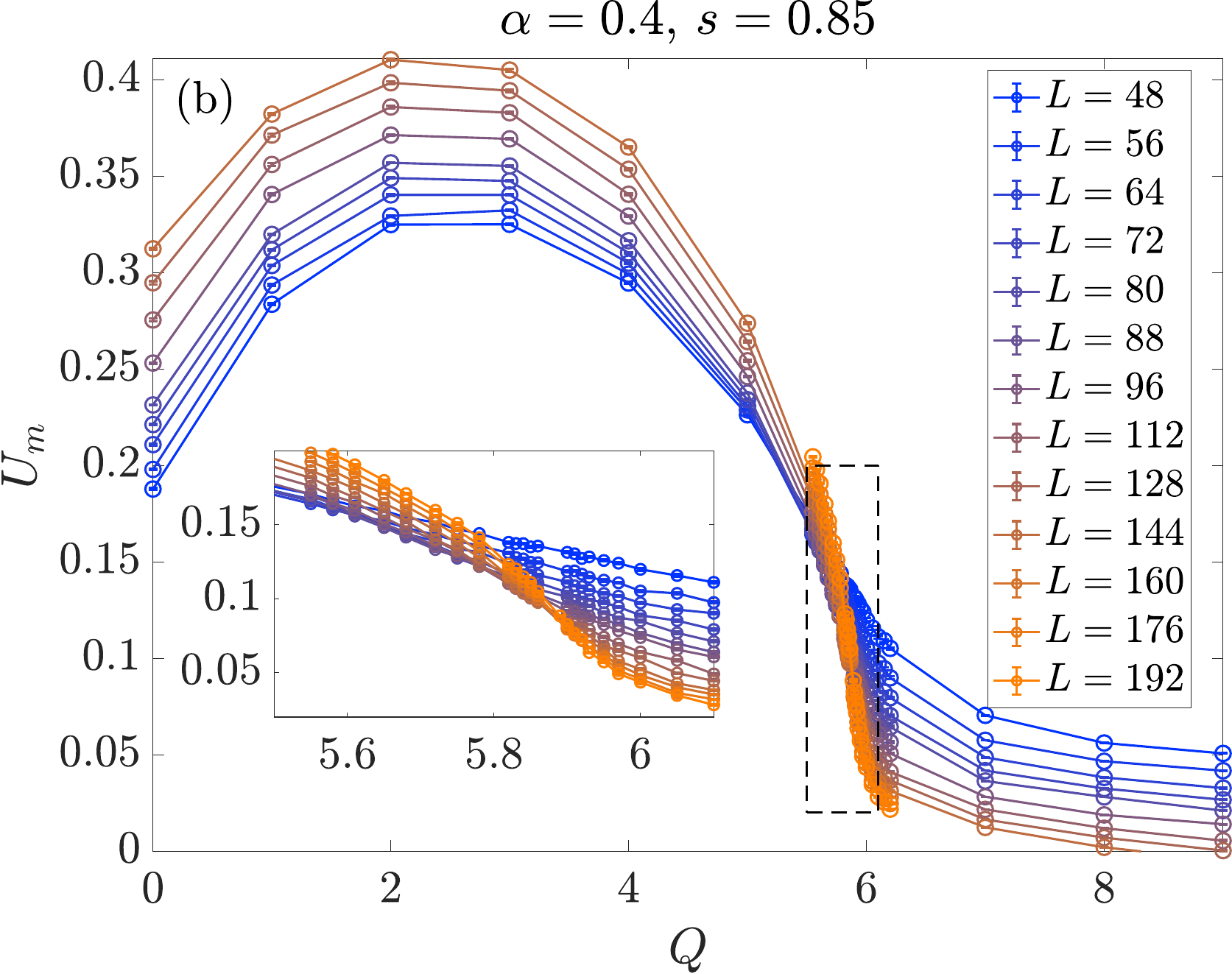}
    \includegraphics[width=0.30\textwidth]{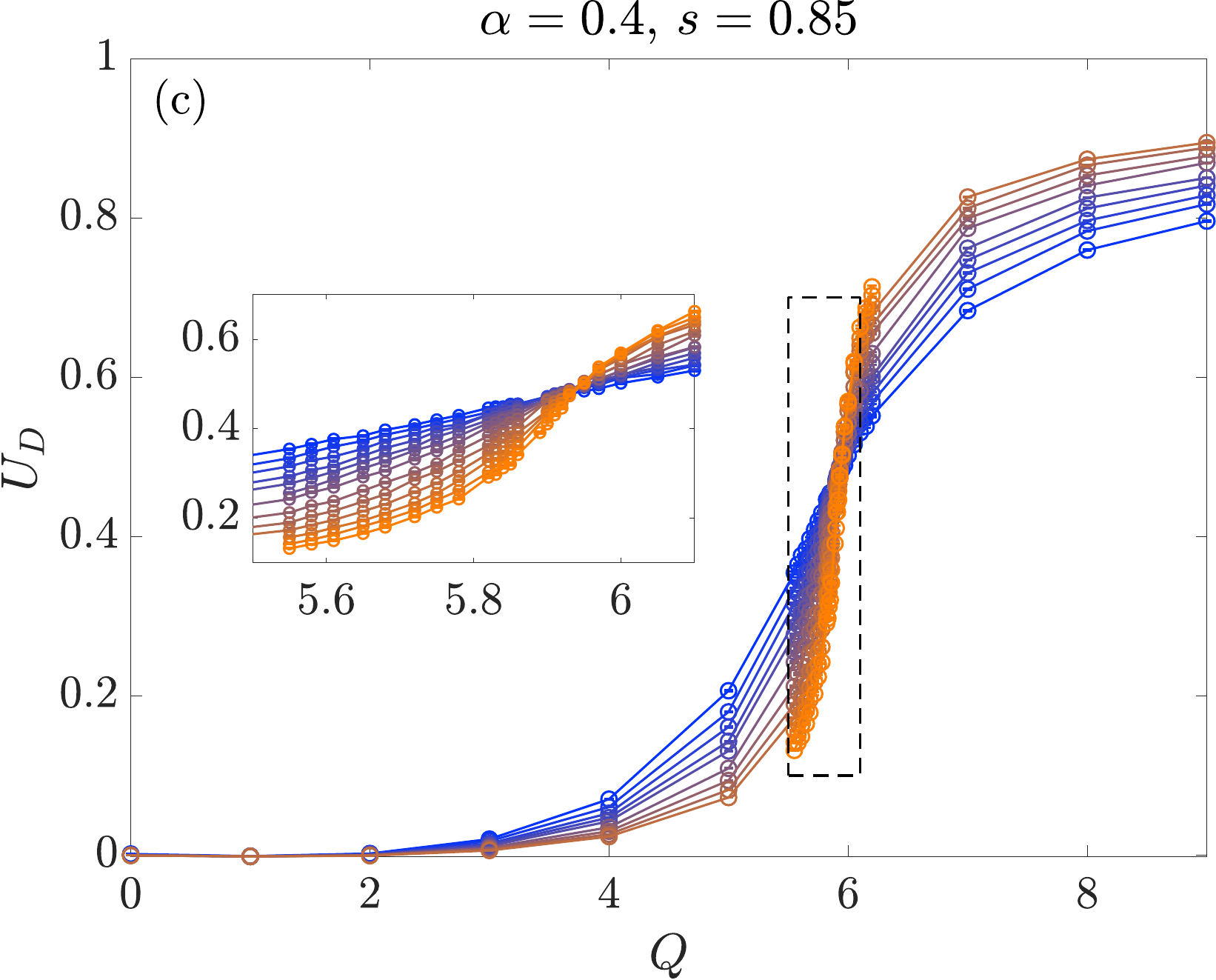}
    \caption{(a) $R$, (b) $U_m$, and (c) $U_D$ versus $Q$ for various system sizes at $\alpha=0.4$ and $s=0.85$. 
    }
    \label{Fig:Binders_0.85}
\end{figure}

Figure~\ref{Fig:Binders_0.85} shows the correlation ratio $R$ and the Binder cumulants $U_m$ and $U_D$ 
as functions of $Q$ for various system sizes at $s=0.85$ and $\alpha=0.4$. 
The simulations were performed with $\beta=0.05L^2$. 
Although for $Q \approx 0$ the temperatures are not sufficiently low to reach the ground state, they are sufficient for $Q \ge 4$, 
enabling us to reliably determine the AFM-VBS transition and its critical properties.

\begin{figure}[htbp]
    \centering
    \includegraphics[width=0.35\textwidth]{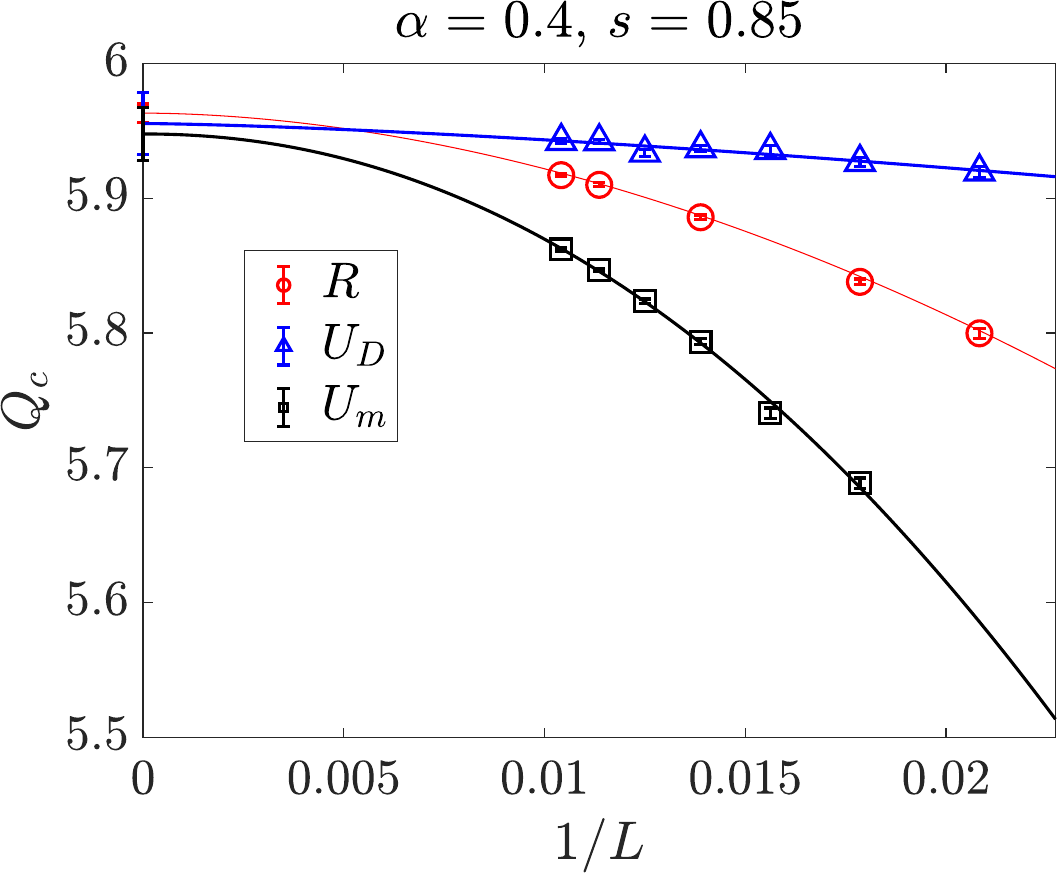}
    \caption{Crossing points analysis for the sub-Ohmic case $s=0.85$ with $\alpha=0.4$. The symbols mark the crossings of the $U_m, R$, and $U_D$ curves between $L$ and $2L$, obtained by polynomial fitting the corresponding curves. The solid lines represent power-law fitting to crossings of $U_m$ and $R$, and cubic polynomial fit to crossings of $U_D$. The extrapolated $Q_c$ are indicated at $1/L=0$.}
    \label{Fig:Qc_s0.85}
\end{figure}

Similar to the Ohmic case presented in the main text, we observe AFM order present at small $Q$ and VBS order at large $Q$. 
The crossings of $U_m(Q)$, $R(Q)$, and $U_D(Q)$ for different system sizes occur at approximately the same $Q$, indicating a direct AFM-VBS transition around $Q\approx 5.7$. 
The crossings $Q_c^{A}(L)$ versus $1/L$ for $A=R, U_m,$ and $U_D$ are graphed in Fig.~\ref{Fig:Qc_s0.85} and converge to the critical point $Q_c=5.963(6)$, as listed in Table I of the main text.

Again, we observe the enhancement of AFM order by $Q$ interactions, which is understood via the 
spin-wave analysis and explained by the RG flow, where the back-scattering $\lambda$ associated with the $Q$ interactions plays an important role. 

\begin{figure}[htbp]
    \centering
    \includegraphics[width=0.60\textwidth]{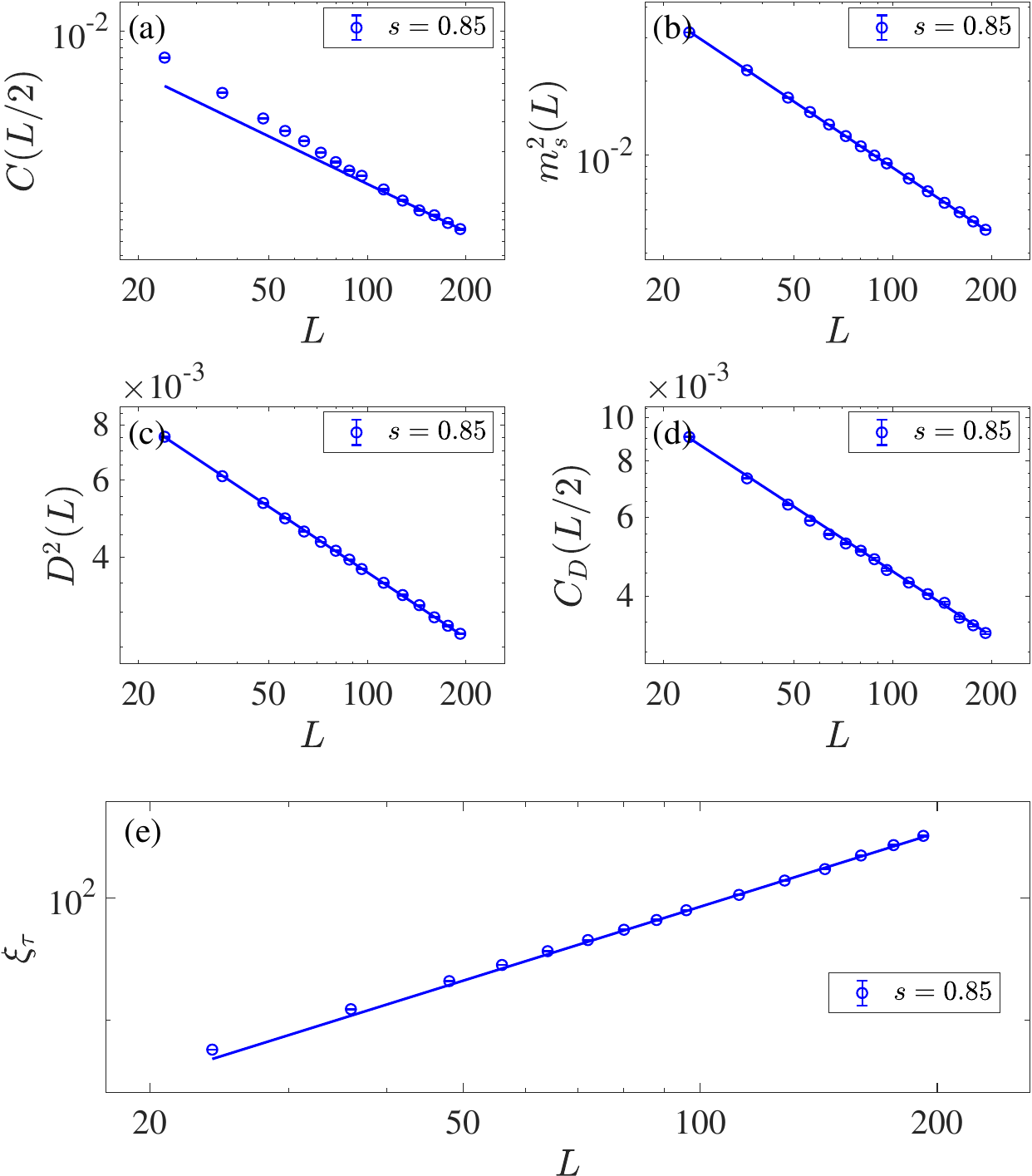}
    \caption{Log-log plots of $C(L/2)$ (a), $m_s^2(L)$ (b), $D^2(L)$ (c),$C_D(L/2)$ (d), and $\xi_\tau$ (e) at 
    the AFM-VBS critical point at $s=0.85$ and $\alpha=0.4$. The error bars are smaller than the symbols. Solid lines are power-law fits.
}
    \label{Fig:cl_s0.85}
\end{figure}

Finite-size behavior of the correlation $C(L/2)$, the squared AFM order parameter $m_s^2(L)$, the squared VBS order parameter $D^2(L)$, VBS correlation $C_D(L/2)$, and the imaginary time correlation length $\xi_\tau$ at the AFM-VBS critical point for the sub-Ohmic case $s=0.85$ with $\alpha=0.4$ is illustrated in Fig. \ref{Fig:cl_s0.85}(a), (b), (c), (d), and (e) , respectively.
From these results, we find the scaling dimensions $\Delta_N$ and $\Delta_D$, and the dynamic critical exponent $z$, which are listed in Table I of the main text.

\begin{figure}[htbp]
    \centering
    \includegraphics[width=0.4\textwidth]{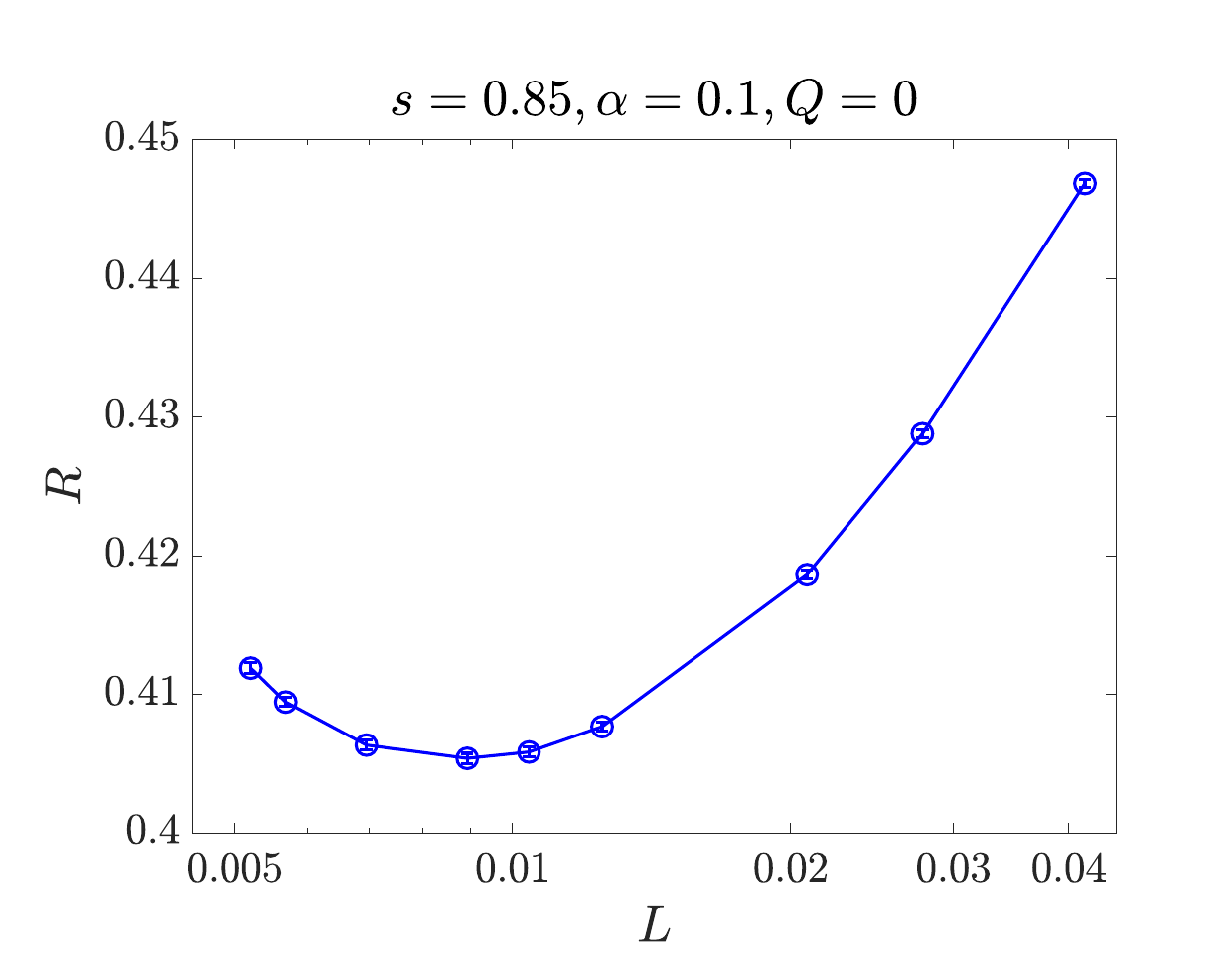}
    \caption{The correlation ratio $R$ versus system size $L$ at $Q=0$, $\alpha=0.4$, and $s=0.85$. A crossover scale $L_c\sim 110$ is clearly seen.} 
    \label{Fig:R_subo}
\end{figure}

Now we present QMC results supporting a crossover from critical state to LRO in the sub-Ohmic case. 
Figure.~\ref{Fig:R_subo}) shows the correlation ratio $R$ as function of system size $L$ at $Q=0, \alpha=0.4$ for sub-Ohmic case $s=0.85$. A crossover length $L_c \sim 110$ is clearly seen. 

\subsection{Violation of $\chi_u$ Scaling at critical point of dissipative model}

\begin{figure}[htbp]
\includegraphics[width=0.30\textwidth]{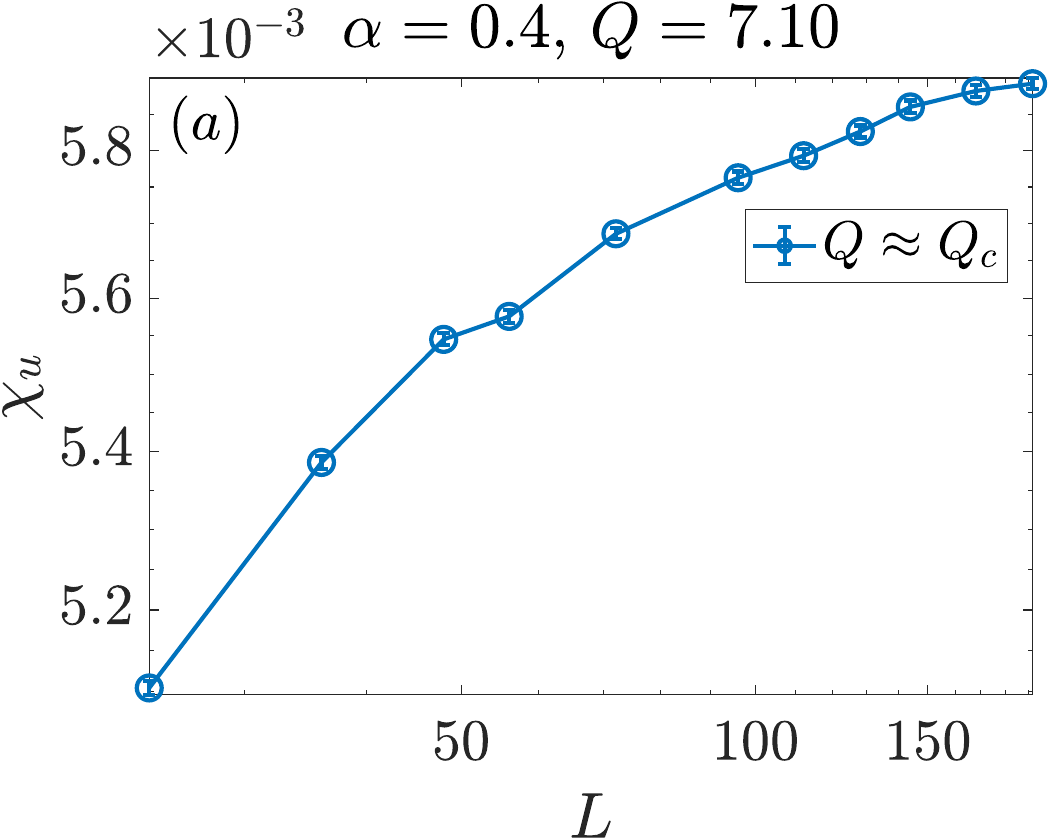}
\includegraphics[width=0.325\textwidth]{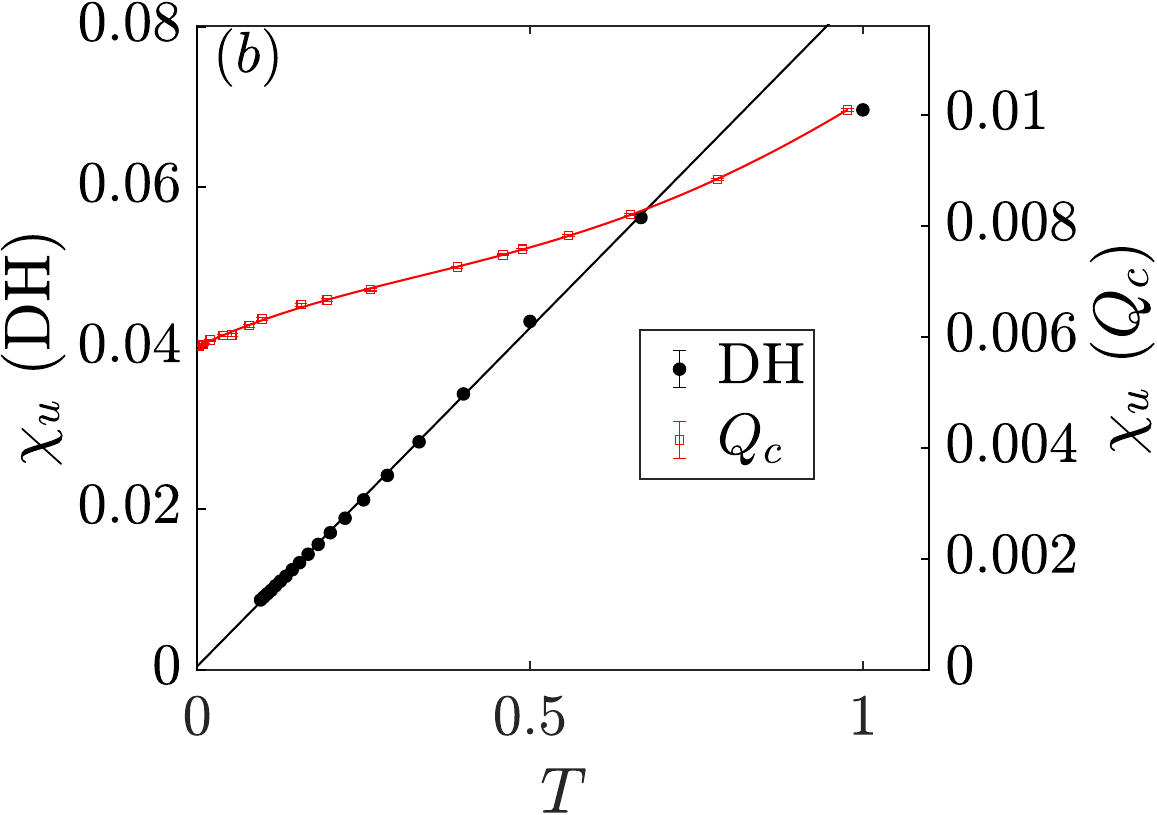}
\caption{(a)The uniform susceptibility $\chi_u$ versus system size $L$ on a log-log scale at the AFM-VBS transition point $Q_c=7.10$ for the Ohmic case. The data were obtained from simulations carried out at inverse temperature $\beta=6L$.
(b) Temperature dependence of $\chi_u$ for the square-lattice columnar dimerized Heisenberg model at the critical point 
(with size $L=64$) and for the Ohmic dissipative $J$-$Q_3$ model at $Q_c=7.10$ (with $L=128$). The solid lines are polynomial fits. For the dimerized Heisenberg model, the temperature dependence is asymptotically linear.
}
\label{Fig:chiu_log}
\end{figure}

The uniform susceptibility is defined as $\chi_u=\beta  \langle M_z^2\rangle/L$, where $M_z=\sum_i S_i^z$ is the total magnetic moment.
At a conventional critical point, $\chi_u$ scales as \cite{Fisher-prb1989, CSY1994}
\begin{equation}
    \chi_u \sim \beta^{1-d/z},
\end{equation}
with the spatial dimension $d=1$ for our model. The corresponding finite-size scaling behavior 
\begin{equation}
    \chi_u(L) \sim L^{z-d}
\end{equation}
is expected. 

However, as shown in Fig. \ref{Fig:chiu_log}(a), $\chi_u$ at the AFM-VBS critical point of the Ohmic dissipation system with $\alpha=0.4$ does not follow such a power law on $L$, indicating that it is not a scaling operator.
This is because the system we study is a subsystem of a larger system-environment composite; therefore, the magnetic moment $M_z$ is no longer a conserved quantity. 
This is demonstrated by the temperature dependence of $\chi_u$ for a finite system size, as shown in Fig. \ref{Fig:chiu_log}(b), where $\chi_u(L)$ converges to a finite value as temperature $T \to 0$.
Such behavior is in sharp contrast to that of isolated AFM systems.
The finite temperature scaling of $\chi_u$ at the critical point of the columnar dimerized Heisenberg model on the square lattice for $L=64$ is also shown in Fig. \ref{Fig:chiu_log}(b) for comparison.

This violation of scaling prevents us from using $\chi_u(L)$ to extract the dynamic critical exponent $z$ at the quantum critical point.

\clearpage

\bibliographystyle{apsrev}  
\bibliography{cs}